\documentclass{emulateapj}

\usepackage{ulem}

\def\xte{{\it RXTE~}} 
\def\js{\object{XTE J1701-462}}

\shortauthors{Lin et al.}

\begin{document}

\title{Spectral States of XTE J1701-462: Link between Z and Atoll Sources}

\author{Dacheng Lin\altaffilmark{1}, Ronald A. Remillard, and Jeroen Homan}
\affil{MIT Kavli Institute for Astrophysics and Space Research, MIT, 70 Vassar Street, Cambridge, MA 02139-4307}
\altaffiltext{1}{email: lindc@mit.edu}

\begin{abstract}

We have analyzed 866 \xte observations of the 2006--2007 outburst of
the accreting neutron star \js, during which the source evolves from
super-Eddington luminosities to quiescence. The X-ray color evolution
first resembles the Cyg X-2 subgroup of Z sources, with frequent
excursions on the horizontal and normal branches (HB/NB). The source
then decays and evolves to the Sco X-1 subgroup, with increasing focus
on the flaring branch (FB) and the lower vertex of the ``Z''. Finally,
the FB subsides, and the source transforms into an atoll source, with
the lower vertex evolving to the atoll soft state. Spectral analyses
suggest that the atoll stage is characterized by a constant inner disk
radius, while the Z stages exhibit a luminosity-dependent expansion of
the inner disk, which we interpret as effects related to the local
Eddington limit. Contrary to the view that the mass accretion rate
($\dot{m}$) changes along the Z, we find that changes in $\dot{m}$ are
instead responsible for the secular evolution of the Z and the
subclasses.  Motion along the Z branches appears to be caused by three
different mechanisms that may operate at roughly constant
$\dot{m}$. For the Sco X-1-like Z stage, we find that the FB is an
instability track that proceeds off the lower vertex when the inner
disk radius shrinks from the value set by the X-ray luminosity toward
the value measured for the atoll soft state. Excursions up the NB
occur when the apparent size of the boundary layer increases while the
disk exhibits little change. The HB is associated with Comptonization
of the disk emission. The Z branches for the Cyg X-2-like stage are
more complicated, and their origin is unclear. Finally, our spectral
results lead us to hypothesize that the lower and upper Z vertices
correspond to a standard thin disk and a slim disk, respectively.

\end{abstract}


\clearpage

\section{INTRODUCTION}
\label{sec:intro}

Based on their X-ray spectral and timing properties, the luminous and
weakly magnetized neutron stars (NSs) in low-mass X-ray binaries
(LMXBs) are classified into atoll and Z sources, named after the
patterns that they trace out in X-ray color-color diagrams (CDs) or
hardness-intensity diagrams (HIDs) \citep{hava,va2006}. Z sources
typically radiate at luminosities close to Eddington luminosity
($L_{\mathrm{EDD}}$), and they trace out roughly Z-shaped tracks in
CDs/HIDs within a few days. Atoll sources cover a lower and larger
luminosity range ($\sim$0.001--0.5 $L_{\mathrm{EDD}}$), and they trace
out their patterns in CDs/HIDs on longer timescales (days to
weeks). {Although extensive coverage by the {\it Rossi X-ray Timing
Explorer} (\xte) has shown that atoll patterns can have Z-like shapes
\citep{murech2002, gido2002a}, they are different from the Z-source
tracks in shape, color ranges, and evolution timescales. Furthermore,
the spectra of Z sources are very soft on all three branches of the
``Z'', whereas the spectra of atoll sources are soft at high
luminosities, but hard when they are faint.  Properties like the rapid
X-ray variability and the order in which the branches are traced out
are also different for the two classes
\citep{baol2002,vavame2003,revava2004,va2006}. The upper, diagonal and
lower branches of the Z-shaped tracks for Z sources are called
horizontal, normal and flaring branches (HB/NB/FB), respectively,
while for atoll sources, they are called the extreme island, island,
and banana states. To stay consistent with our previous work on atoll
sources, however, we refer to the atoll branches as hard, transitional
and soft states (HS/TS/SS), respectively.

\begin{figure}
\plotone{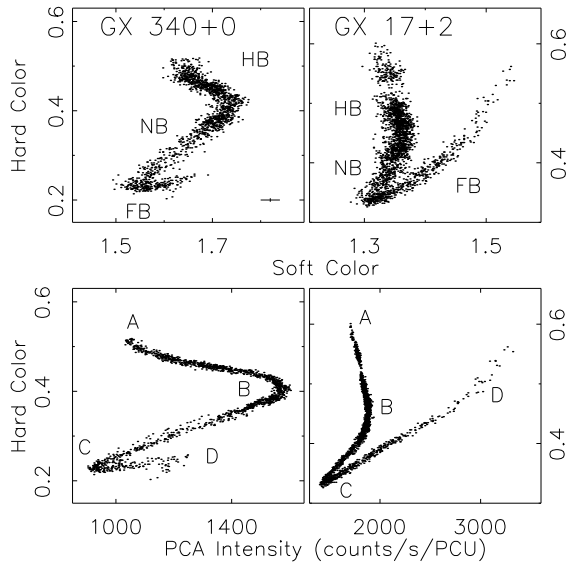} 
\caption{CDs and HIDs of the Cyg-like Z source GX~340+0 (MJD
51920--51925) and the Sco-like Z source GX~17+2 (MJD 51454--51464),
with bin size 128 s. The typical error bars for the colors are shown in
the top left panel; errors in the intensity are
negligible. The Z-source branches (HB, NB, and FB) are labeled in the
CDs. 'A', 'B', 'C', and 'D' in the HIDs mark key positions in the Z
tracks: the left end of HB, HB/NB vertex, NB/FB vertex, and the right
end of FB, respectively. Their corresponding spectra are shown in
Figure~\ref{fig:twopha_pcahexgx340gx17}.
\label{fig:ccdiaggx340gx17}} 
\end{figure}

\begin{figure*}\epsscale{1.0}
\plotone{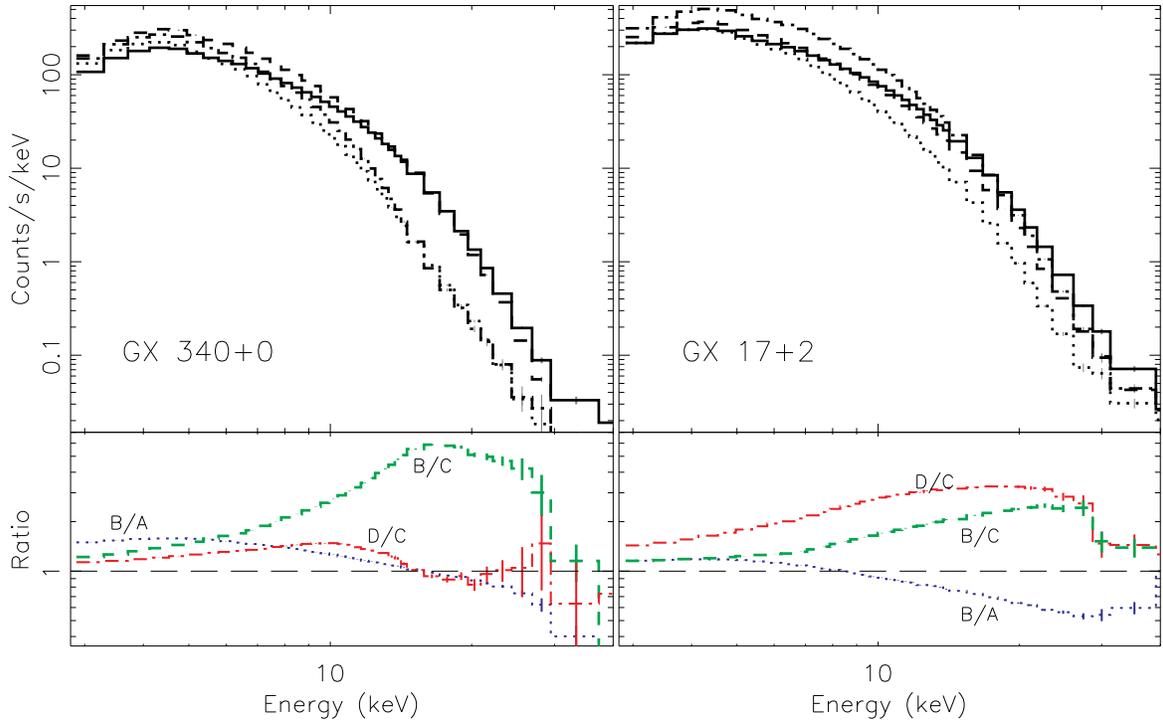} 
\caption{PCA spectra of GX~340+0 and GX~17+2 from key positions along
their Z tracks. The solid, dashed, dotted, and dot-dashed lines
correspond to labels 'A'--'D' in Figure~\ref{fig:ccdiaggx340gx17}
respectively. The ratios of the spectra from the two ends of each
branch are shown in the bottom panels. Spectra with high total PCA
intensity are divided by those with lower total PCA intensity in order
to show that on each branch the intensity increases in a different
energy range.
\label{fig:twopha_pcahexgx340gx17}} 
\end{figure*}

\begin{figure*} \epsscale{1.0}
\plotone{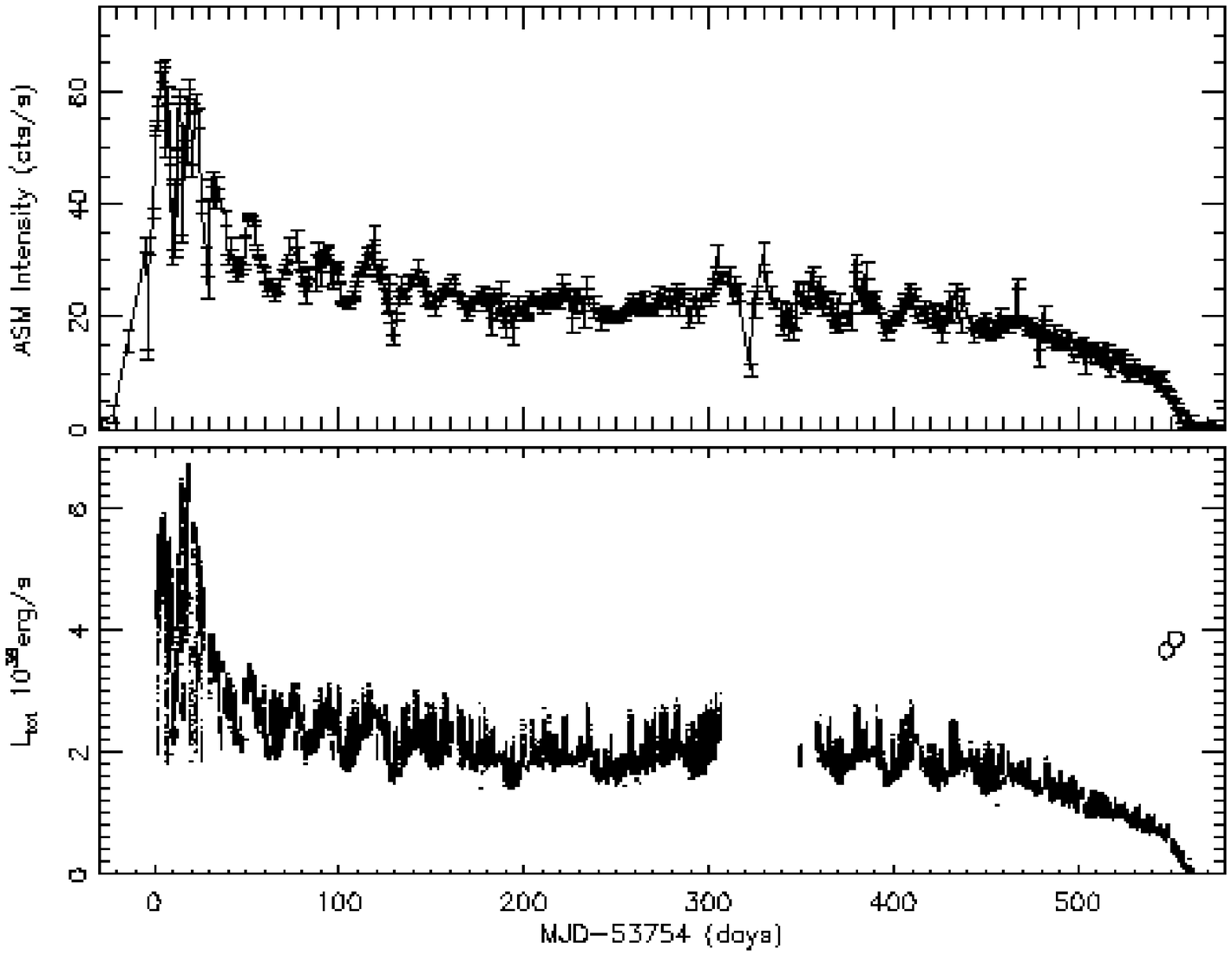}
\caption{Upper panel: the \xte ASM one-day-averaged light curve of
XTE~J1701--462 during its 2006--2007 outburst; lower panel: the \xte
PCA 32-s luminosity curve, from spectral fits with a MCD+BB model and
assuming the source distance to be 8.8 kpc and system inclination
70$\degr$ (\S\ref{sec:specmod}). The two circles mark the peak luminosities (persistent
emission subtracted) of two type I X-ray bursts, detected from this
source during the decay of the outburst. These two bursts showed
photospheric radius expansion, thus indicating the Eddington limit.
\label{fig:asmdata}} 
\end{figure*}

Based on the shape and orientation of their branches, the six
classical Z sources were further divided into two subgroups
\citep{kuvaoo1994}: Cyg-like (\object{Cyg X-2}, \object{GX 340+0}, and
\object{GX 5-1}) and Sco-like (\object{Sco X-1}, \object{GX 17+2}, and
\object{GX 349+2}). We show sample CDs and HIDs for these subgroups in
Figure~\ref{fig:ccdiaggx340gx17}. The spectra from key positions along
the ``Z" are shown in Figure~\ref{fig:twopha_pcahexgx340gx17}. We also
plot the ratios of these spectra at the two ends of each branch
(bottom panels) to show that the motion along each Z branch is the
result of spectral changes in different energy ranges. The spectral
differences between the two subgroups are quite apparent, especially
in the case of the FB (red dot-dashed lines in
Figure~\ref{fig:twopha_pcahexgx340gx17}). Although the branches have
the same names for each subgroup, their origins are possibly
different. In addition to movement along the ``Z" tracks, the Z tracks
themselves display slow shifts and shape changes in CDs/HIDs. These
so-called secular changes are most apparent in \object{Cyg X-2}.

There are several questions regarding the Z sources that remain
unanswered: e.g., what is the nature of the Z branches, how do they
relate to the spectral states of atoll sources, and how are the two Z
subclasses related? A unique opportunity to improve our understanding
of Z sources arose with the discovery in 2006 of \js\
\citep{reli2006}, the first NS transient to show all the
characteristics of a Z source \citep{hovawi2007}. In the first 10
weeks of its $\sim$600-day outburst, \js\ transformed from a Cyg-like
into a Sco-like Z source \citep{hovawi2007}, and during the decay it
evolved further into an atoll source \citep{howial2007}.  The upper
and lower panels of Figure~\ref{fig:asmdata} show light curves of the
outburst, using data from, respectively, the All-Sky Monitor
\citep[ASM;][]{lebrcu1996} and the Proportional Counter Array
\citep[PCA;][]{brrosw1993, jaswgi1996} on board \xte. The latter one
shows the luminosity as obtained from spectral fits (see
\S\ref{sec:specmod} for more details). The two circles correspond to
the peak luminosities of two type I X-ray bursts (persistent emission
subtracted) which showed photospheric radius expansion
\citep{lihore2007,lihoal2009}. In terms of a single value of Eddington
limit (however, see \S\ref{sec:conclusion}), we see from this figure
that the source reached super-Eddington luminosities during the peak
of its outburst, assuming orbital inclination to be $70\degr$.

The large dynamic range in luminosity of \js, from super-Eddington
down to near-quiescence, also implies significant changes in the mass
accretion rate ($\dot{m}$). This allows one to investigate the
relevance of $\dot{m}$ to the questions that we posed above. The study
by \citet{hovawi2007} suggests that differences in $\dot{m}$ are
responsible for the Z-source subclasses, with the Cyg-like sources
accreting at higher rates. Initial results from the end phase of the
outburst suggest that the differences between Z and atoll sources are
also purely the results of a difference in $\dot{m}$, with a lower
$\dot{m}$ for the atoll class \citep{howial2007}.

Concerning the role of $\dot{m}$ in the evolution along the Z tracks,
we note that results from multi-wavelength campaigns have been
interpreted as monotonically increases in $\dot{m}$ from the HB,
through the NB, to the FB
\citep[e.g.,][]{havaeb1990,vrraga1990}. However, this classical view
is inconsistent with the fact that the X-ray intensity decreases as Z
sources move along the NB in the direction of the FB. In recent years,
several alternatives have been proposed. Based on the behavior of the
0.1--200 keV flux as reported by \citet{distro2000} and on a
comparison with black hole systems, \citet{hovajo2002} suggested that
$\dot{m}$ might be constant along the Z track, with motion along the
``Z" being caused by an unknown parameter. \citet{chbaja2008} used
X-ray spectral fits to claim that $\dot{m}$ increases in the direction
opposite to the classical view, being lowest on the FB, which is a
branch proposed to be driven by unstable nuclear burning.  Finally,
based on the observed change between Cyg-like and Sco-like Z-source
behavior in the NS transient \js, \cite{hovawi2007} proposed
that secular changes in Z sources are the result of changes in
$\dot{m}$ and that the position along the ``Z" is determined by
$\dot{m}$ normalized by its long term average. Definitive conclusions
on the role of $\dot{m}$ have been hindered by the lack of a spectral
model that lets us unambiguously track the evolution of physical
components along the Z track.

A major difficulty in interpreting the X-ray spectra of NS LMXBs has
been the problem of model degeneracy, i.e., significantly different
models providing acceptable fits to the same data \citep[][hereafter
LRH07]{lireho2007}. In the classical framework of two-component models
for the continuum spectra of NS LMXBs, one component is thermal, and
the other is Comptonized. The thermal component can be either a
single-temperature blackbody (BB), used to describe the boundary
layer, or a multicolor disk blackbody (MCD), while there are choices
for the nature of the Comptonized component, with no
clear advantages for any set of combination of these two
components. As far as Z sources are concerned, \citet{dozysm2002},
\citet{agsr2003}, and \citet{dazydi2007} used a model of
Comptonization plus a MCD to fit the spectra of \object{Cyg X-2},
\object{GX 349+2} and \object{Sco X-1}, respectively, while
\citet{distro2000,diroia2001,difabu2002} used a model of
Comptonization plus a BB for \object{GX 17+2}, \object{GX 349+2} and
\object{Cyg X-2}.  The model used by \citet{chbaja2008} for their
study of \object{GX 340+0} also consisted of a BB plus Comptonization,
with the latter approximated by a cut-off power law.

In LRH07, we showed for two atoll type transients that the commonly
used spectral models for thermal emission plus Comptonization are not
favored for the SS, because they fail to satisfy various desirability
criteria, such as $L_{\rm X} \propto T^4$ evolution for the MCD
component and similarity to black holes for correlated timing/spectral
behavior. In an attempt to solve this, we devised a hybrid model: a BB
to describe the boundary layer plus a broken power-law (BPL) for the
HS, and two strong thermal components (MCD and BB) plus a constrained
BPL (when needed) for the SS. The results from this model are
summarized as follows: both the MCD and BB evolve as $L_{\rm X}
\propto T^4$, the spectral/timing correlations of these NSs are
aligned with the properties of accreting black holes, and the visible
BB emission area is very small but roughly constant over a wide range
of $L_{\rm X}$ that spans both the HS and SS. We note that this model
is still partially empirical, especially the (constrained) BPL
description of Comptonization. We also note that the boundary layer
spectrum was reported to be broader than a BB in the Z-source HB and
atoll-source SS \citep{giremo2003,regi2006}.  However, it is also
possible that the broadening is caused by rapid variability of a
Comptonized component and its blending with a BB in these
states/branches.

Considering the success of this spectral model for atoll sources, we
intend to apply it to \js. This unique source was observed on an
almost daily basis with \xte\ during its outburst. In this work, we
present a color/spectral analysis of all \xte\ observations from the
2006--2007 outburst of \js, as part of our campaign to understand this
source (\citealt{lihoal2009} (type I X-ray bursts);
\citealt{hoetal2009} (transition from a Z to an atoll source);
\citealt{fretal2009} (quiescence)). The goal of this paper is to
improve our understanding of the physical processes that drive the
evolution along the Z-source tracks and the link between the Z-source
branches and atoll-source states. A description of our data set and
reduction techniques is given in \S\ref{sec:reduction}. In
\S\ref{sec:colordiag}, we present an analysis of the CDs/HIDs. States
and branches are classified for the entire outburst, and we study the
global evolution of the Z/atoll tracks. Using our state/branch
classification, in \S\ref{sec:specmod} we present our spectral fit
results as a function of increasing source luminosity. Each of the
Z-source branches and transitions between them (i.e., the vertices) is
examined in terms of repeatable patterns in the evolution of spectral
parameters. Results of a brief investigation of timing properties are
presented in \S\ref{sec:timing}. In \S\ref{sec:discussion}, we present
a physical interpretation of the evolution of the source and some of
the branches and vertices of the Z track, and we discuss the role of
$\dot{m}$. A comparison with other NS LMXBs is also given. Finally, we
summarize our main conclusions in \S\ref{sec:conclusion}.

\section{OBSERVATIONS AND DATA REDUCTION}
\label{sec:reduction}

We analyzed all 866 pointed observations ($\sim$3 Ms) of \js\ from the
2006--2007 outburst made with the PCA and the High Energy X-ray Timing
Experiment \citep[HEXTE;][]{roblgr1998} instruments. For the PCA we
only used data from Proportional Counter Unit (PCU) 2, which is the
best-calibrated unit. For the HEXTE only Cluster B data were used,
because all the observations of this outburst were made after 2006
January, when Cluster A started experiencing frequent problems with
rocking between the source and background positions. We used the same
standard criteria to filter the data as described in LRH07. Only
observations with PCA intensity (background subtracted) larger than 10
counts/s/PCU were considered. Appropriate faint/bright background
models were used when the source had intensity lower/higher than 40
counts/s/PCU. Deadtime corrections for PCA and HEXTE data were also
applied.

To spectrally model the evolution of the source during the entire
outburst, we must create pulse-height spectra on an appropriate
timescale. This timescale should be short enough to characterize the
spectral evolution, but also long enough to minimize statistical
uncertainties. \object{XTE J1701-462} showed rapid spectral variations
on some occasions, especially on the FB. To characterize these
variations, we created spectra with exposures of 32 s and 960 s, from
``standard 2'' data of PCU 2. The 960-s spectra were created by
combining 30 sequential 32-s spectra with a maximum observational gap
$<$500 s. Although we allow several gaps in a 960-s spectrum, in the
end all such spectra have time span $<$1600 s. Longer observation gaps
than 500 s limit the exposure to $<$960 s. If a spectrum had an
exposure $<$320 s, it was combined with the preceding spectrum if the
time gap between them was $<$500 s; otherwise it was omitted. The
final selection of spectra for modeling was based on our understanding
of the source variability, using the 32-s and 960-s spectra (see \S
\ref{sec:variability}).

Once we had spectra from the PCA for spectral modeling, we created the
response file for each spectrum using HEAsoft version 6.4. An
investigation of Crab Nebula observations revealed that the low energy
channels still showed calibration issues after 2006. Therefore, we
used channels 4--50 (numbering from 1 to 129; $\sim$2.7--23.0 keV) for
our spectral fitting and applied systematic errors of $0.8\%$ for
channels 5--39 (below 18 keV) and $2\%$ for channels 4 and 40--50. We
created the HEXTE spectra to match the PCA spectra in observation
time. No systematic errors were applied for HEXTE data.

The observations of \js\ were contaminated with a small amount of
diffuse Galactic emission within the large field of view of the PCA
(${\sim}1\degr$) \citep{fretal2009}. Count rates near the end of the
outburst, between Modified Julian Day (MJD) 54321 and 54342, reached a
constant level, ${\sim}2$ counts/s/PCU (after the normal background
subtraction), due to this diffuse component. The next \xte pointed
observation of this source was nearly 5 months later and indicated a
similar level of residual flux. To correct for the diffuse emission,
we created a spectrum from the observations between MJD 54321 and
54342 and applied it as an extra background correction. We only did
this for observations between MJD 54300 and 54321, when the source
intensity was $<300$ counts/s/PCU and the diffuse emission could skew
the spectral results significantly.

In order to study the source evolution in a model-independent manner,
we also examined the CDs/HIDs of \js. We calculated X-ray colors as
described in LRH07. We normalized the raw count rates from PCU 2 using
observations of the Crab Nebula, and we defined soft and hard colors
as the ratios of the normalized background-subtracted count rates in
the (3.6--5.0)/(2.2--3.6) keV bands and the (8.6--18.0)/(5.0--8.6) keV
bands, respectively. Throughout the paper, all PCA intensity values
are normalized and correspond to the sum count rate of these four
energy bands. The difference between the normalized intensity and the
raw total PCA count rate is normally $<$$5\%$.

\section{LIGHT CURVES AND COLOR-COLOR DIAGRAMS}
\label{sec:colordiag}

\subsection{Source State/Branch Classification}

\begin{figure*}\epsscale{1.0}
\plotone{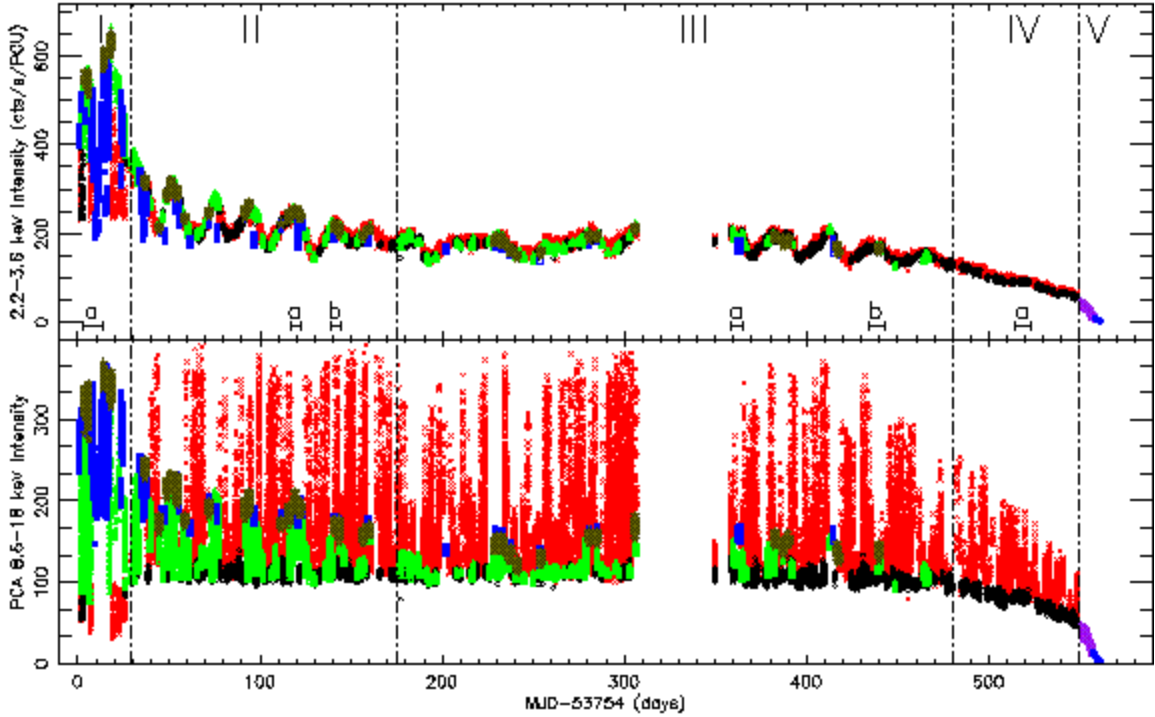} 
\caption{\xte PCA 32-s light curves of XTE~J1701--462 in two energy
bands during the 2006--2007 outburst. The typical error bars are
smaller than the symbol size. The outburst is divided into five
stages. In stages I--IV, the source showed characteristics of Z
source, with the HB, NB and FB marked by blue squares, green
triangles, red crosses, respectively. We also mark the HB/NB and the
NB/FB vertices by olive pentagrams and black diamonds, respectively
(Figure~\ref{fig:sumhid}). In stage V, the source showed
characteristics of atoll source, with the HS and SS marked by blue
filled circles and purple hexagrams, respectively. The 'a' and 'b' in each stage
mark the sample intervals for which detailed source properties are
shown in \S\ref{sec:sampleintervals1} and \S\ref{sec:zmodelfit}.
\label{fig:lctwobands}}
\end{figure*}

\begin{figure}
\plotone{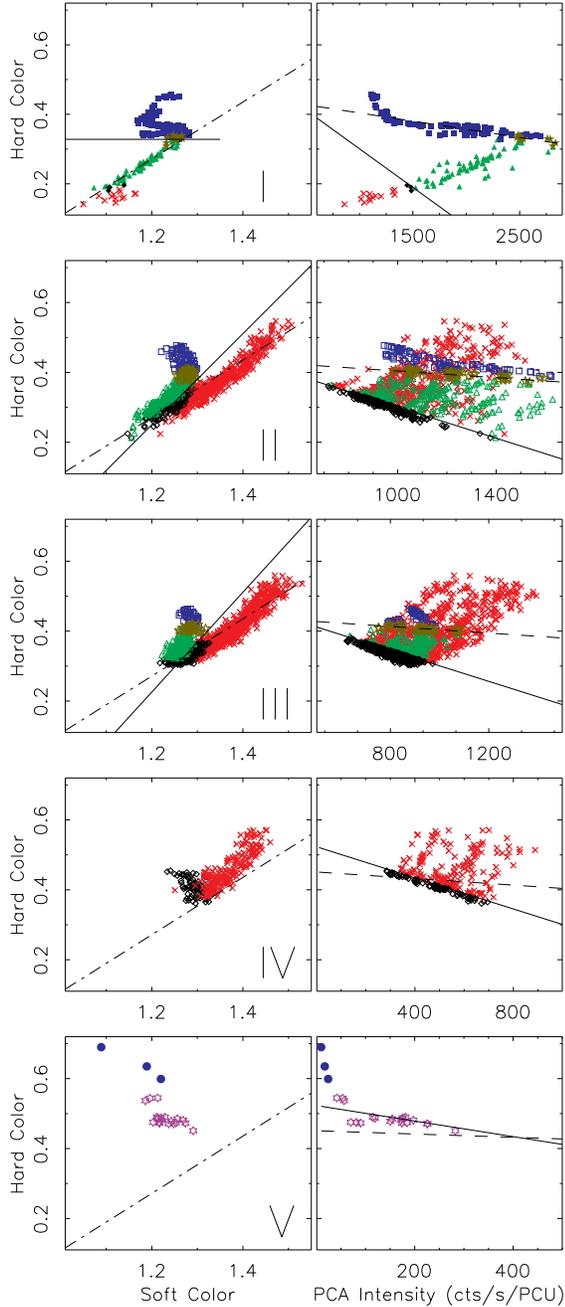}
\caption{The CDs and HIDs for the five stages of the outburst defined
in Figure~\ref{fig:lctwobands}. The bin size is $\sim$960 s for stages
I--IV, while for stage V each data point corresponds either to one PCA
observation or several observations combined (when the count rates are
$<$30 counts$/$s$/$PCU). The meaning of the symbols are the same as in
Figures~\ref{fig:lctwobands} and \ref{fig:sumhid}. The solid and
dashed lines are used to define the source branches. The typical error
bars are smaller than the symbol size. The dot-dashed line is aligned
with the NB in the CD of stage I and is shown in other panels of CDs
for reference .
\label{fig:ccdiagfivestages}}
\end{figure}

\begin{figure*} \epsscale{1.0}
\plotone{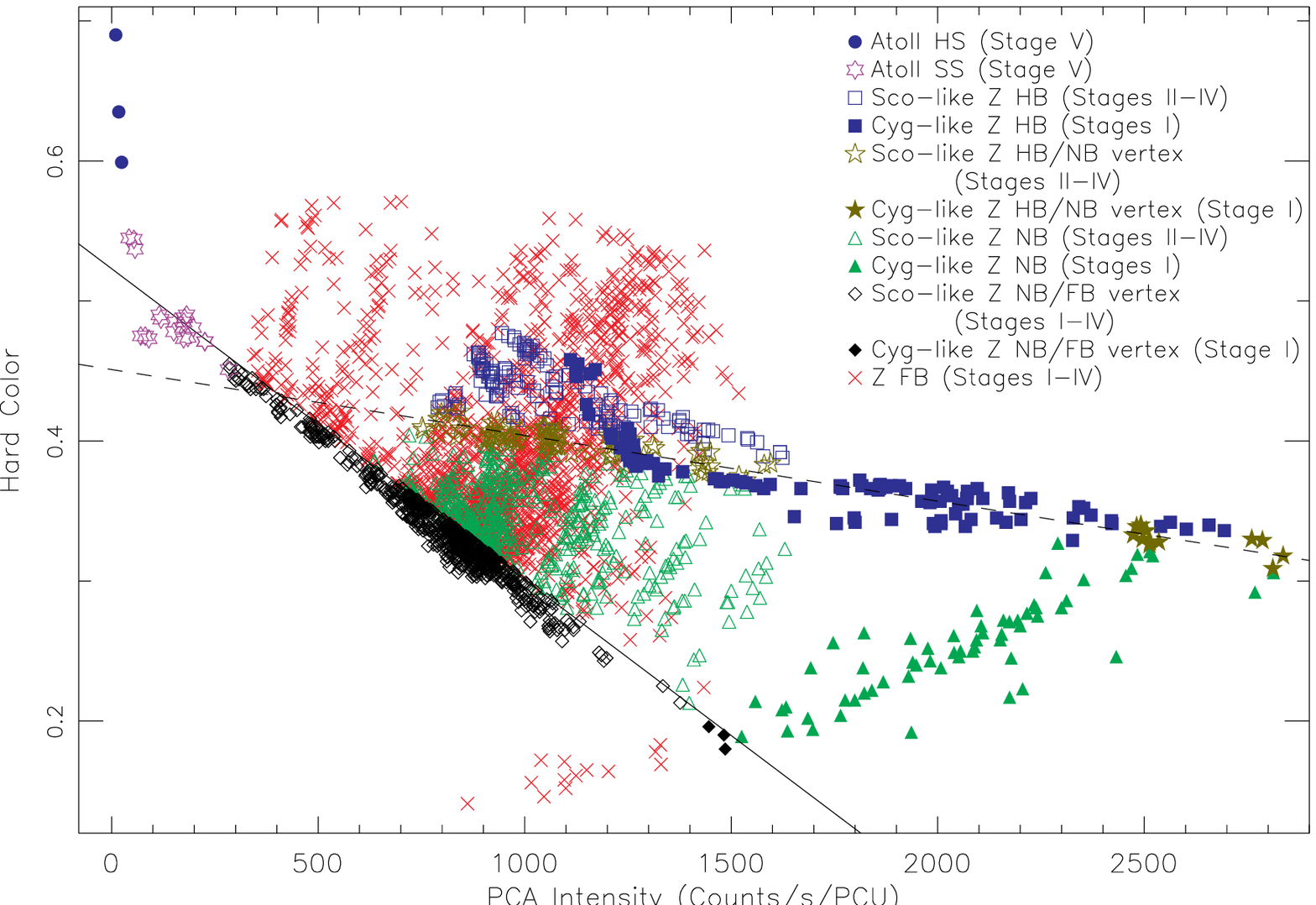}
\caption{The complete HID of the outburst of XTE~J1701--462. For
 reference, a legend of all different symbols is shown in the upper
 right corner; they apply to all other figures in this paper.
\label{fig:sumhid}}
\end{figure*}

Normalized light curves, CDs and HIDs of the outburst of \js\ are
shown in Figures~\ref{fig:lctwobands} and
\ref{fig:ccdiagfivestages}. The two light curves in
Figure~\ref{fig:lctwobands} have bin size 32 s, and they show count
rates from the 2.2--3.6 and 8.6--18.0 keV energy bands,
respectively. They are quite different, with the high-energy curve
showing rapid and strong flaring, and the low-energy curve showing a
$\sim$25-day modulation \citep{hovawi2007} that varies in strength
with time. The CDs and HIDs in Figure~\ref{fig:ccdiagfivestages} were
made from 960-s spectra to minimize the statistical uncertainties. The
entire outburst is divided into five time stages, denoted by Roman
numerals I-V. Their boundaries are marked by vertical dot-dashed lines
in Figure~\ref{fig:lctwobands}, and the CDs and HIDs corresponding to
each of the stages can be found in
Figure~\ref{fig:ccdiagfivestages}. The boundaries between the stages
are somewhat arbitrary, but are based on the following considerations:
(I/II, MJD 53783) the source switches from Cyg-like to Sco-like Z
source behavior in the CD, as reported in \citet{hovawi2007}; (II/III,
MJD 53929) the long-term modulations at low energies become weaker and
slower; (III/IV, MJD 54232) the long-term modulations disappear, and
the final decay starts; (IV/V, MJD 54303) the flaring at high energies
ends.

Most of the observations in stages I--II were analyzed in detail by
\citet{hovawi2007} and shown to exhibit Z-source
characteristics. Stage I is characteristic of a Cyg-like Z source, and
stages II--III are similar to the Sco-like Z sources
(Figure~\ref{fig:ccdiagfivestages}). In stage IV the source is similar
to some of the bright atoll sources, such as \object{GX 9+1} and
\object{GX 9+9} \citep{hobewi2007}. While in those sources the
observed flaring would be referred to as an upper banana branch, our
analysis (see below) shows that it is simply a lower luminosity
version of the flaring branch. In our view, the flaring is one of the
defining characteristics that separate the Sco-like Z sources from the
atoll sources. Thus we group stage IV with stages I--III and refer to
them as the Z (source) stages. In stage V, \js\ showed characteristics
of an atoll source \citep{howial2007}, and we will refer to this stage
as the atoll (source) stage. Finally, we note that the secular changes
in stages II and III are mainly the result of the long-term
modulations.

An important step toward understanding the evolution of the source is
the classification of the observations in terms of states and
branches. For sources showing substantial secular changes, this is
normally done by dividing the data sets into smaller subsets, in order
to identify clear tracks in the CDs/HIDs
\citep[e.g.,][]{wivaku1997,hovawi2007}. The observations of
\object{XTE J1701-462} are not long and dense enough (typically one
hour per day) to do this for the entire data set, as the secular
changes often smear the Z tracks before they are completely covered. A
different way of classification is therefore necessary. Fortunately,
we can take advantage of the fact that each branch shows systematic
evolution in the CDs/HIDs of each of the stages. The HID for the
entire outburst is shown in Figure~\ref{fig:sumhid}; the position of
the Z vertices throughout the outburst is marked by the solid and
dashed lines, and the plot symbols that we use to differentiate the
states/branches are given for reference (see below). We note that
clear vertex lines are still seen if we define colors using the
different energy bands that were used in \citet{hovawi2007}.

The atoll stage V is divided into the HS (blue filled circle, with HC $>$ 0.55)
and SS (purple hexagram). Since little short-time variation is seen in
stage V, each data point in Figure~\ref{fig:ccdiagfivestages}
corresponds to one entire PCA observation, and we also combine several
observations when the source intensity is $<$30 counts/s/PCU. These
same data intervals for stage V are used for spectral analysis in
\S\ref{sec:specmod}.

The branch classification strategy for Z stages proceeds as
follows. Stages II--IV are Sco-like, and we focus on them first. In
these stages, the HB, NB, FB, HB/NB vertex and NB/FB vertex are marked
by (open) blue square, green triangle and red cross, olive pentagram
and black diamond symbols, respectively (Figure~\ref{fig:sumhid}). We
made a preliminary branch classification in the CDs by hand, using
data in $\sim$10-day time intervals, and found that the NB/FB vertex
always hovers near the bottom of the corresponding HID. During the
outburst the motion of this vertex is described well by a single line
(i.e., the solid line) in the HID of Figure~\ref{fig:sumhid}. Data
below this line are identified as NB/FB vertex in stages II--IV. This
line is also shown in the HIDs of stages I and V in
Figure~\ref{fig:ccdiagfivestages} for reference. The FB in stages
II--IV is most easily identified in the CDs; it can be separated from
other branches by simple lines (solid lines in the CDs of
Figure~\ref{fig:ccdiagfivestages} for stages II--III), while excluding
the NB/FB vertex points described above. In stage IV, the HB and NB
are absent, and the FB consists of all the points that remain after
the NB/FB vertex points are identified.

For stages II and III, the HB, NB and HB/NB vertex classifications
still need to be defined. We first manually identified HB/NB vertex
points based on the CDs and HIDs for a few short intervals (${\sim}10$
days). Again, we found that they lie near a single line in the HIDs,
i.e., the dashed line in Figure~\ref{fig:sumhid}. This dashed line
is also shown for reference in the HIDs of
Figure~\ref{fig:ccdiagfivestages}. We identify the observations with a
HC value within 0.01 of this line as the HB/NB vertex, excluding those
on the FB, which were identified earlier. The other observations above
and below this line were identified as the HB and NB respectively. At
this point, the branch classification for Sco-like Z stages II--IV is
complete. The classification process can be summarized as follows: the
NB/FB vertex is identified by a line near the bottom of HIDs, the FB
is separated from other branches in the CDs, and the HB/NB vertex lies
around a single line in the HIDs, above and below which are the HB and
NB respectively (excluding the FB and NB/FB vertex).

Cyg-like stage I spans about one month and mainly consists of two
tracks. It shows strong dips (lower panel in
Figure~\ref{fig:lctwobands}). The solid line which limits the NB/FB
vertex in stages II--IV in the HIDs also applies to this stage
(Figure~\ref{fig:ccdiagfivestages}), with the `dipping' FB falling
below the line, rather than above it. The boundary between the HB and
NB is defined by a constant HC line in the CD. The HB/NB vertex points
for the two main tracks are identified by hand. They fall close to the
extension of the vertex line found for stages II--IV, as shown in
Figure~\ref{fig:sumhid} (dashed line). The HB, NB, FB, HB/NB vertex
and NB/FB vertex in this stage are marked by (filled) blue square,
green triangle and red cross, olive pentagram and black diamond
symbols, respectively (Figure~\ref{fig:sumhid}).

\subsection{Source Evolution and Relations between Source Types}
\label{sec:evolution}

\begin{deluxetable*}{cccccc}
\tabletypesize{\scriptsize}
\tablecaption{Statistics for observation time of the source in different states/branches\label{tbl-1}}
\tablewidth{0pt}
\tablehead{
&  \multicolumn{5}{c}{observation time in ks and (percentage of the total)}\\
\cline{2-6} \\
\colhead{Source state/branch} &  \colhead{I} & \colhead{II} &  \colhead{III} &\colhead{IV} &\colhead{V}
}
\startdata
HB           & 96 (55)    & {\ }90 (9){\ }  &  40 (3)  & 0 (0) & ---\\
HB/NB vertex & 10 (6){\ } & {\ }42 (4){\ } &  33 (3)  & 0 (0) & ---\\
NB           & 52 (31)    & 228 (24) & 178 (14) & 0 (0) & ---\\
NB/FB vertex & 2 (1)      & 201 (21) & 291 (24) & 77 (33) & ---\\
FB           & 12 (7){\ } & 408 (42) & 687 (56) & 156 (67){\ } & ---\\

SS& --- & --- & ---&---&54 (58)\\
HS & --- & --- & ---&---&40 (42)\\
\enddata
\end{deluxetable*}

With the classification of the source states/branches in place, we now
investigate the global evolution of \object{XTE J1701-462} during this
outburst. Table~\ref{tbl-1} gives a summary of the amount of time that
the source is observed in the different branches/states of each
stage. Although these values are somewhat dependent on the definition
of the boundaries between the source states/branches, it is quite
clear that the fractions of time that the source stays on the HB and
NB decrease with the evolution of the outburst with that on the HB
decreasing faster. The source stays on the NB/FB vertex and the FB
more and more frequently from stage I to stage IV. Moreover, while the
source spends most of the time ($>$$90\%$) on the HB and NB in stage
I, it is not observed on the HB and NB at all in stage IV. We note
that during each stage, the source often moves back and forth along
the NB or the FB without entering another branch
(Figure~\ref{fig:lctwobands}).  Although there are no discontinuous
jumps from one branch to another, the source often reverses its
direction within a branch. We do not see a clear constraint as to when
a source enters a specific branch, which seems to be random. The only
exception is the occurrence of the HB during stages II and III, which
appears mostly, but not exclusively, around the peaks of the long term
modulation as observed in the low-energy light curve
(Figure~\ref{fig:lctwobands}).

Following the disappearance of the HB and NB during stage IV, the FB
disappears in stage V, and the original NB/FB vertex smoothly
evolves into an atoll track. This is reflected both in the light
curves (Figure~\ref{fig:lctwobands}) and in the HID
(Figure~\ref{fig:sumhid}). Moreover, there is no observational
evidence that suggests that the HB evolves into the atoll SS or the
atoll HS as $\dot{m}$ decreases \citep{murech2002,
gido2002a,hovawi2007}. Another interesting phenomenon seen in
Figure~\ref{fig:sumhid} is that the distance between the HB/NB vertex
and the NB/FB vertex in the HID decreases with intensity (or
equivalently, the NB shortens as the intensity decreases). In the HID,
the HB/NB vertex line crosses that of the NB/FB vertex near the point
where the FB disappears and the atoll track starts.

The Cyg-like HB/NB and NB/FB vertices seem to be natural extensions of
Sco-like ones at high intensity (Figure~\ref{fig:sumhid}). The
Cyg-like NB also resembles the Sco-like ones. However, there are clear
differences between the Cyg-like dipping FB and Sco-like strong
FB. The HB in the Cyg-like stage is also very distinctive. It spans a
much larger range of intensity than in other stages
(Figure~\ref{fig:sumhid}). However, in contrast with the FB, whose
pointing direction switches from downward to upward in the HID as soon
as the source enters stage II, the HB seems to transit more slowly, as
it is still nearly horizontal and has significant range of intensity in
stage II, especially at high intensity
(Figure~\ref{fig:ccdiagfivestages}). This can be seen more clearly
from sample intervals in \S\ref{sec:sampleintervals1}. The HB becomes
vertical only when the intensity is $\sim$800-1300 counts/s/PCU
(Figure~\ref{fig:sumhid}). It is interesting to see in
Figure~\ref{fig:sumhid} that the nearly horizontal part of the
Cyg-like HB (excluding the upturn) is closely aligned with the line
traced out by the HB/NB vertex. This might hint at some connection
between that part of the Cyg-like HB and the HB/NB vertex in the
Sco-like Z stages. The upturn of the HB of the Cyg-like stage happens
to have an intensity similar to the Sco-like HB. Considering that they
are both vertical in the HID, they might be related, which we
investigate based on spectral and timing properties later in this
work.

\subsection{Evolution Speed along Branches}
\label{sec:speed}

\begin{figure}
\plotone{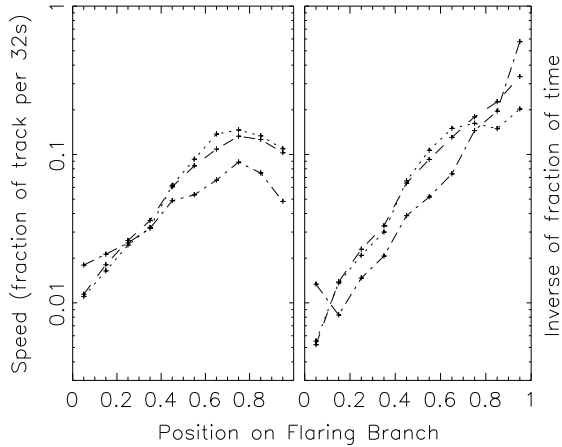}
\caption{Left panel: the speed of the source along the FB, expressed
as the average fraction of the FB track that the source goes through
in 32 s. The FB is normalized to
have length one, with the NB/FB vertex set to be 0.0. The position on
the FB is based on the 8.6-18.0 keV intensity. The statistical error
is very small, but the sample standard deviation is about as large as
the speed itself. Right panel: inverse of the fraction of time in each
segment (the normalization is arbitrary). This quantity is an
alternative measure of the speed of the source along the FB
(\S\ref{sec:speed}).
\label{fig:colorspeedFB}}
\end{figure}

\begin{figure}
\plotone{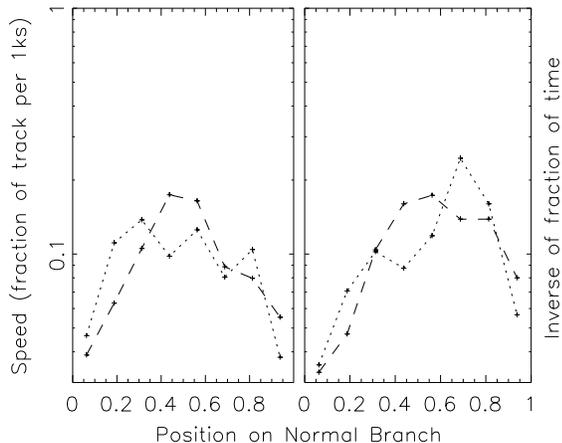}
\caption{Same as Figure~\ref{fig:colorspeedFB} but for the NB. The
NB/FB vertex has a position value of 0.0, and the HB/NB vertex is 1.0. Time bins of 960 s are used for this plot.
\label{fig:colorspeedNB}}
\end{figure}

Clues regarding the physical origin of each Z branch might be
reflected in the timescales on which the source evolves along it. We
therefore measure the speed at which \js\ moves along its Z track.
Such speed measurements have been made in the past for other Z sources
\citep[e.g.,][]{wivaku1997,hovajo2002}. These authors used a rank
number to track the source position along the Z track, which is scaled
to the full length of the NB. This is very difficult to implement for
\object{XTE J1701-462}, as it shows large secular changes of the
tracks in the CDs/HIDs, making it hard to universally assign a rank
number for all tracks. Instead, we measure the speed for each branch
separately, by choosing a quantity $X$ that changes substantially and
is suitable for tracing the position along that branch. The normalized
position in that branch can then be expressed as $S=(X-X_{\rm
min})/(X_{\rm max}-X_{\rm min})$, where the maximum ($X_{\rm max}$)
and minimum ($X_{\rm min}$) values are obtained in a time-dependent
manner. When sequential data points have position numbers $S_i$ and
$S_{i+1}$ and centroid temporal separation $\Delta{t}$, the speed at
position $(S_i+S_{i+1})/2$ is calculated as
$|S_{i+1}-S_i|/\Delta{t}$. This method is hereafter referred to as the
rank-shift method. Using the normalized position in a branch, we also
calculate a second measure of evolution speed, which is the inverse
value of the fraction of the time that the source spends at each
position of the branch. This method is valid under the assumption that
our sampling does not bias the results.

For the FB and the NB/FB vertex, our speed measurement utilizes the
intensity in the energy band 8.6-18.0 keV, since it shows strong
changes (Figure~\ref{fig:lctwobands}). The source moves very fast
along the FB, and 32-s spectra are used. We find that the intensity in
this band maintains nearly constant minimum ($\sim$100 counts/s/PCU)
and maximum levels ($\sim$350 counts/s/PCU) during stages II--III. Thus
we set $X_{\rm min}=100$ counts/s/PCU and $X_{\rm max}=350$
counts/s/PCU for stages II--III. For stage IV, the range of the
intensity at 8.6-18.0 keV on the FB changes with time. Thus we divide
stage IV into intervals of 20 days and dynamically define $X_{\rm
min}$ and $X_{\rm max}$ for each interval. Stage I has few FB
observations, which are all dipping; they are not investigated.

The rank-shift results for the FB are shown in the left panel of
Figure~\ref{fig:colorspeedFB}, using dotted, dashed and dot-dashed
lines for stages II, III, and IV respectively. The NB/FB vertex has
position number 0.0. We average the speed for ten bins along the
FB. The statistical errors are very small, but the sample standard
deviations per bin have values of order the speed. It is clear that
the speed increases by about an order of magnitude as the source
ascends the FB. The plot shows a slight decrease in speed at the tip
of the FB, but this may be an artifact of the source reversing its
direction at this position; if the speed at the top of the FB is high,
this can occur within the timescale of our 32-s measurements. Based
on the measured speed, we estimate that it takes $\sim$10 minutes for
\object{XTE J1701-462} to cross the FB in one direction. The right
panel of Figure~\ref{fig:colorspeedFB} is the inverse of the fraction
of the time that \object{XTE J1701-462} spends at each position of the
FB. This result resembles the rank-shift method, in support of our
finding that the speed increases as \object{XTE J1701-462} ascends the
FB.

The situation for the NB is more complicated.
Figure~\ref{fig:lctwobands} suggests that the intensity at 8.6-18.0
keV is also a good tracer for the position along the NB; the NB is
much more pronounced here than in the low energy band. Samples of the
NB show that it takes much longer time, about several hours, for
\object{XTE J1701-462} to travel across the NB. Hence, we use 960-s
spectra to reduce the effect of short-time fluctuations. We only
select intervals with complete tracks of the NB, i.e., including both
the HB/NB and NB/FB vertices. Only stages II and III are considered,
since only these two stages provide sufficient data. The results are
shown in Figure~\ref{fig:colorspeedNB}. Because of its shorter range,
the NB is divided into eight bins. Again we show the rank-shift method
on the left and the method of the inverse fraction of time on the
right. The NB/FB and HB/NB vertices have position numbers 0.0 and 1.0,
respectively. Both stages and both methods give a common result: the
speeds at the HB/NB and NB/FB vertices are much smaller than in the
middle of the NB. Based on the speed measured here, we estimate that
it takes about 3 hours for the source to cross the NB (excluding the
first and last bins, assumed to be occupied by the vertices). The NB
and FB branch crossing times have been independently confirmed from
light curves and CDs/HIDs.

It is harder to measure the speed along the HB. Data samples
(\S\ref{sec:sampleintervals1}) show that it takes $\sim$a day for the
source to trace out the HB, a timescale at which the secular changes
of Z tracks become significant. The intensity decreases at high
energy, but increases at low energy as the source climbs up the HB
from the HB/NB vertex (\S\ref{sec:sampleintervals1}). This suggests
that a broad hardness ratio like (2.2--3.6 keV)/(8.6-18.0 keV) can be
an effective means to track the source position on the HB. However,
even this quantity only changes by at most $20\%$ in stages
II--III. Moreover, the end point of the HB is not well defined, which
impedes our efforts to measure the evolution speed along the HB.

In summary, for the Sco-like Z branches it takes $\sim$10 minutes,
$\sim$3 hours, and $\sim$a day for the source to go through the FB,
NB, and HB respectively. The NB/FB vertex appears to represent a point
of increased stability, compared to the adjoining branches, since the
source evolves much slower when it enters this
vertex. Figure~\ref{fig:colorspeedNB} shows that the source slows down
as it enters the HB/NB vertex from the NB, but we do not know whether
the evolution speeds up as the source leaves the HB/NB vertex along
the HB. Figures~\ref{fig:colorspeedFB} and \ref{fig:colorspeedNB} can
be interpreted to suggest that the FB and NB are unstable
configurations relative to their ending vertices. We cannot say much
about the Cyg-like Z stage I, due to insufficient observations, but a
few samples also shows that the source moves much slower on the HB
than on the NB or FB.

\subsection{Selection of Spectra for Spectral Fits}

In this section, we examine the variability of \object{XTE J1701-462}
in different energy bands. The variability results will guide the data
selection for our spectral fits, to be conducted over the entire
outburst. We also investigate in detail the properties of a few sample
intervals that exhibit evolution along Z branches. These samples will
also be used later for our spectral analysis, to help us understand
the physical causes of evolution along each branch.

\subsubsection{Variability in Different Energy Bands}
\label{sec:variability}

\begin{figure*} \epsscale{1.0}
\plotone{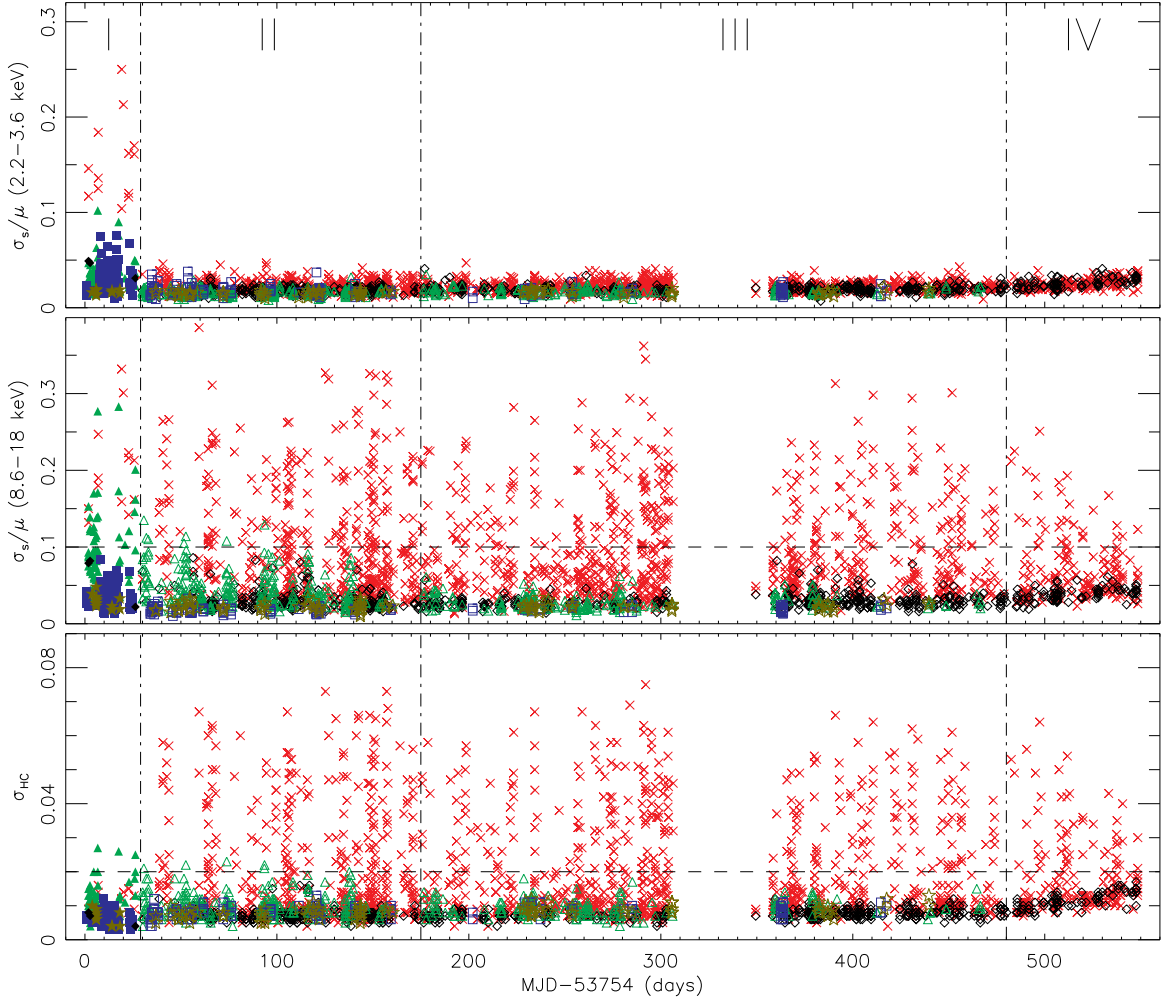}
\caption{Different measures of spectral variability during the Z
stages. Each point corresponds to one 960-s spectrum. The top two
panels show the fraction of variability in two energy bands, using
quanta of 32-s spectra for each point. The bottom panel is the
corresponding sample standard deviation of the hard color.
\label{fig:color_stat}}
\end{figure*}

The spectra in the atoll stage V show little variation within a
typical PCA observation (Figure~\ref{fig:lctwobands}), and the
spectrum of complete observations can be used for spectral
fitting. However, variations are more significant and faster during
the Z stages I--IV. We quantify the variations of the spectra within
the time span of each 960-s spectrum using the 32-s spectra that it
contains (see \S \ref{sec:reduction}). We calculate the sample
standard deviations ($\sigma_s$) and the means ($\mu$) of the source
intensity in the low and high energy bands and plot the fraction of
variability ($\sigma_s/\mu$) in Figure~\ref{fig:color_stat}. In the
low energy band the intensity varies little (${\sim}2\%$) within
${\sim}960$ s in all branches, except for stage I where it varies up
to ${\sim}25\%$ on the dipping FB. 

In the high energy band, the source intensity on the FB can vary up to
30$\%$ within ${\sim}960$ s, while the HB and the NB typically show
${<}10\%$ variability. Also plotted in Figure~\ref{fig:color_stat} is
the sample standard deviation of the hard color ($\sigma_{\rm HC}$),
again using the quanta of 32-s spectra within a given 960-s
spectrum. Its variability is similar to the source intensity at high
energy: $\sigma_{\rm HC}$ is ${<}0.02$ (${\sim}4\%$) except on the FB
where it can be up to $0.08$ (${\sim}15\%$).

For a 960-s spectrum we adopt the following criteria for steady
conditions: $\sigma_{\rm HC} \leq 0.02$ and $\sigma_{\rm s}/\mu\ ({\rm
8.6-18.0 keV}) \leq 10\%$. The steady 960-s spectra generally have
excellent statistics for spectral modeling. We obtain 2374 such
spectra in the Z stages, about 79$\%$ of the total. Almost all
non-steady intervals are on the FB.

\subsubsection{Sample Intervals}
\label{sec:sampleintervals1}

\begin{figure*} \epsscale{1.0}
\plotone{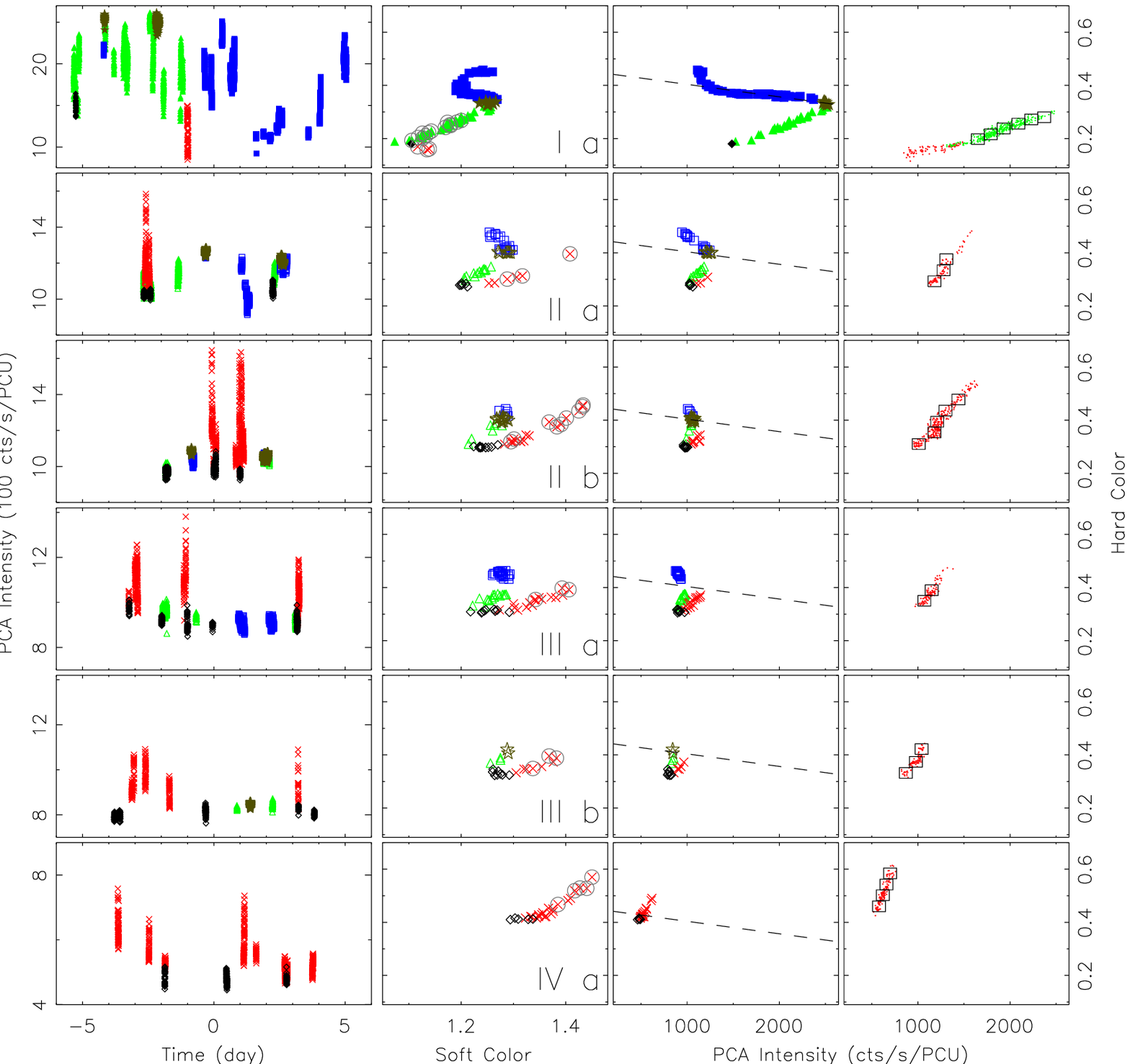}
\caption{Light curves, CDs and HIDs of the sample intervals marked in
Figure~\ref{fig:lctwobands}. The light curves are from 32-s spectra
and the CDs from $\sim{960}$-s spectra. The HIDs for the steady
(\S\ref{sec:colordiag}) and non-steady spectra (circled in the CDs)
are shown in the third and fourth columns, respectively. The non-steady
HIDs use the 32-s subintervals, and spectra within each box shown in
these HIDs are combined to form a spectrum used for spectral fitting
in \S\ref{sec:zmodelfit}.
\label{fig:colors_sub}}
\end{figure*}

\begin{figure}
\plotone{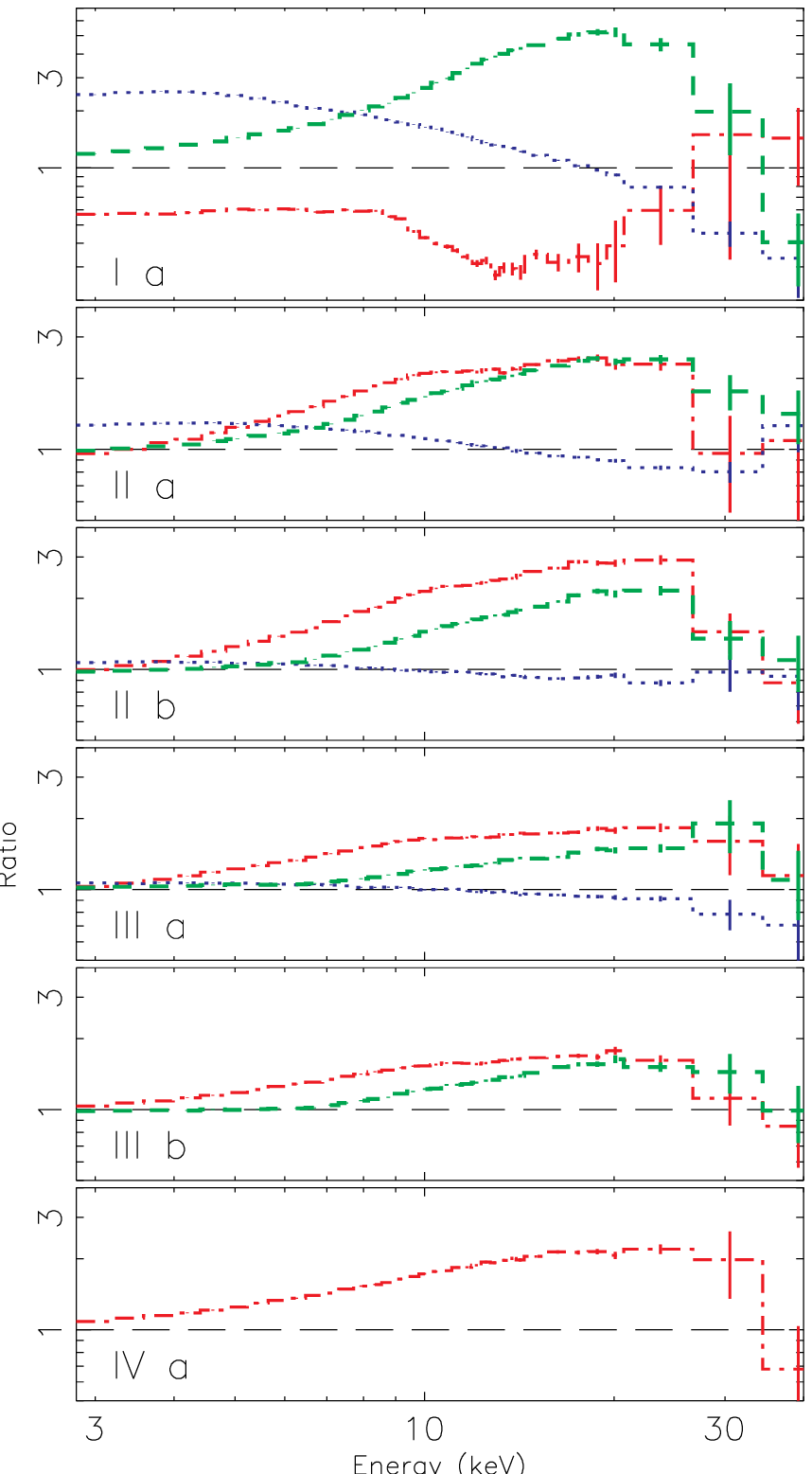} 
\caption{Comparison of the PCA spectra of XTE~J1701-462 in
key positions along the Z tracks from the sample intervals
(\S\ref{sec:sampleintervals1}). Blue dotted lines represent the ratio of the
spectra at the HB/NB vertex relative to the left end of the HB. Green
dashed lines show the ratio of the spectra at the HB/NB vertex
relative to the NB/FB vertex. Red dot-dashed lines show the ratio of
the spectra at the outer end of FB relative to the NB/FB vertex.
\label{fig:twopha_pcahexx1701}}
\end{figure}

We now examine several sample Z tracks from the time intervals marked
in Figure~\ref{fig:lctwobands}, corresponding to MJD 53756.6--53767.3,
53870.5--53876.2, 53893.0--53898.0, 54112.0--54119.0,
54188.0--54196.5, 54268.5--54277.0, and 54290.0--54299.0,
respectively. These intervals span about five to ten days and are
selected based on their small secular changes in the CDs/HIDs. Their
light curves, CDs and HIDs are shown in Figure~\ref{fig:colors_sub};
the first and second columns are the light curves and CDs,
respectively, and HIDs are plotted in the third and fourth
columns. The light curves have bin size 32 s as in
Figure~\ref{fig:lctwobands}, but in this case the intensity is the sum
of all four energy bands (2.2--18.0 keV). From the light curves, we
see that the source evolves back and forth between different branches
more than once in each sample interval. However, the corresponding CDs
and HIDs (960-s bin size) each show only a single track. We therefore
conclude that the secular changes in each sample interval are
small. For the HIDs we differentiate two types of 960-s spectra. Data
points from steady intervals are plotted in the third column.  The
steady 960-s spectra are used for spectral fitting without further
grouping. We circle the non-steady 960-s spectra in the CDs and show
their corresponding HIDs in the fourth column using their 32-s
subintervals. To gain statistical precision, while avoiding spectral
variability, we use boxes in the HIDs to group similar 32-s spectra
and produce one spectrum per box (with exposures $>$300 s) for
spectral fitting in \S\ref{sec:specmod}.

We can compare the CDs/HIDs of the sample intervals from \js\ with
those from other Z sources (Figure~\ref{fig:ccdiaggx340gx17}). To aid
in this comparison we plot in Figure~\ref{fig:twopha_pcahexx1701} the
ratios of spectra from key positions along the Z tracks in these
sample intervals, analogous to what is displayed in
Figure~\ref{fig:twopha_pcahexgx340gx17}. The blue dotted lines
correspond to the ratio of the spectra at the HB/NB vertex relative
to the open end of the HB, the green dashed lines show the ratio of
the spectra at the HB/NB vertex relative to the NB/FB vertex, and the
red dot-dashed lines show the ratio of the spectra at the outer end
of the FB relative to the NB/FB vertex. The ratios are defined to
divide spectra with higher PCA intensity by ones with lower PCA
intensity. The comparison of
Figures~\ref{fig:colors_sub}--\ref{fig:twopha_pcahexx1701} with
Figures~\ref{fig:ccdiaggx340gx17}--\ref{fig:twopha_pcahexgx340gx17}
confirms that stage I is consistent with being Cyg-like, while stages
II--IV are Sco-like.

Figure~\ref{fig:twopha_pcahexx1701} (blue dotted lines) shows that
when the source descends the HB toward the HB/NB vertex, the intensity
increases at low energies, but decreases at high energies. The pivot
point is ${\sim}20$ keV in interval Ia, decreasing to ${\sim}10$ keV
in the later intervals. We note that the HB in the interval IIa seems
to represent conditions that are still in transition from a Cyg-like
to a Sco-like Z source based on such spectrum ratio. This is
consistent with the fact that the HB in this interval spans a large
intensity range and appears horizontal in the HID, similar to Cyg-like
sources (Figures~\ref{fig:ccdiaggx340gx17}). The increase in the
intensity as the source moves from the NB/FB vertex to the HB/NB
vertex (green dashed line) reaches maximum at energies $>10$
keV. Moreover, in later intervals the intensity is almost constant
below 7 keV. As for the FB, in interval Ia the intensity decreases as
the source evolves along the FB from the NB/FB vertex, signifying a
`dipping' FB \citep{hovawi2007}. This is different from the typical
Cyg-like Z sources
(Figures~\ref{fig:ccdiaggx340gx17}--\ref{fig:twopha_pcahexgx340gx17}),
where intensity initially increases from the NB/FB vertex along the
FB, sometimes followed by dips that appear midway along the FB. The FB
of \js\ in stage I only shows dipping. This dipping increases suddenly
at energies between 10--20 keV (Figure~\ref{fig:twopha_pcahexx1701}),
perhaps indicating some type of occultation effect. In the Sco-like
stages, the FB is quite long in the CDs/HIDs and the source intensity
increases when moving away from the NB/FB vertex, most obviously in
the high energy band 10--30 keV. All these details support the
conclusion that stages II--IV are Sco-like.

\section{SPECTRAL MODELING}

\label{sec:specmod}

\subsection{Spectral Models and Assumptions}

\begin{figure}
\plotone{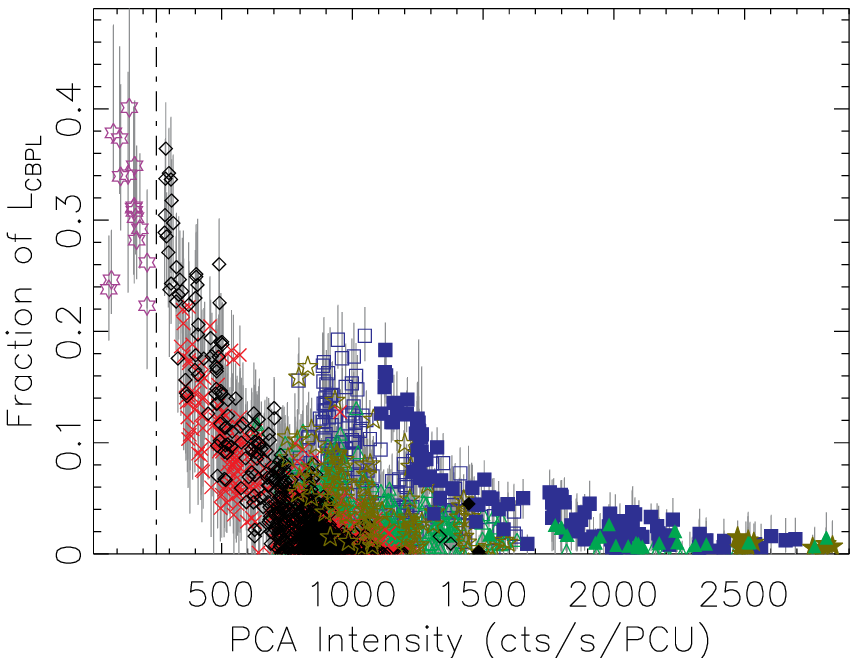} 
\caption{The fraction of the $L_{\rm CBPL}$ with one-$\sigma$ error
bars in the SS of atoll stage and in the steady spectra from all the Z
stages.
\label{fig:xtej1701model5model6com}} 
\end{figure}

\begin{figure}
\plotone{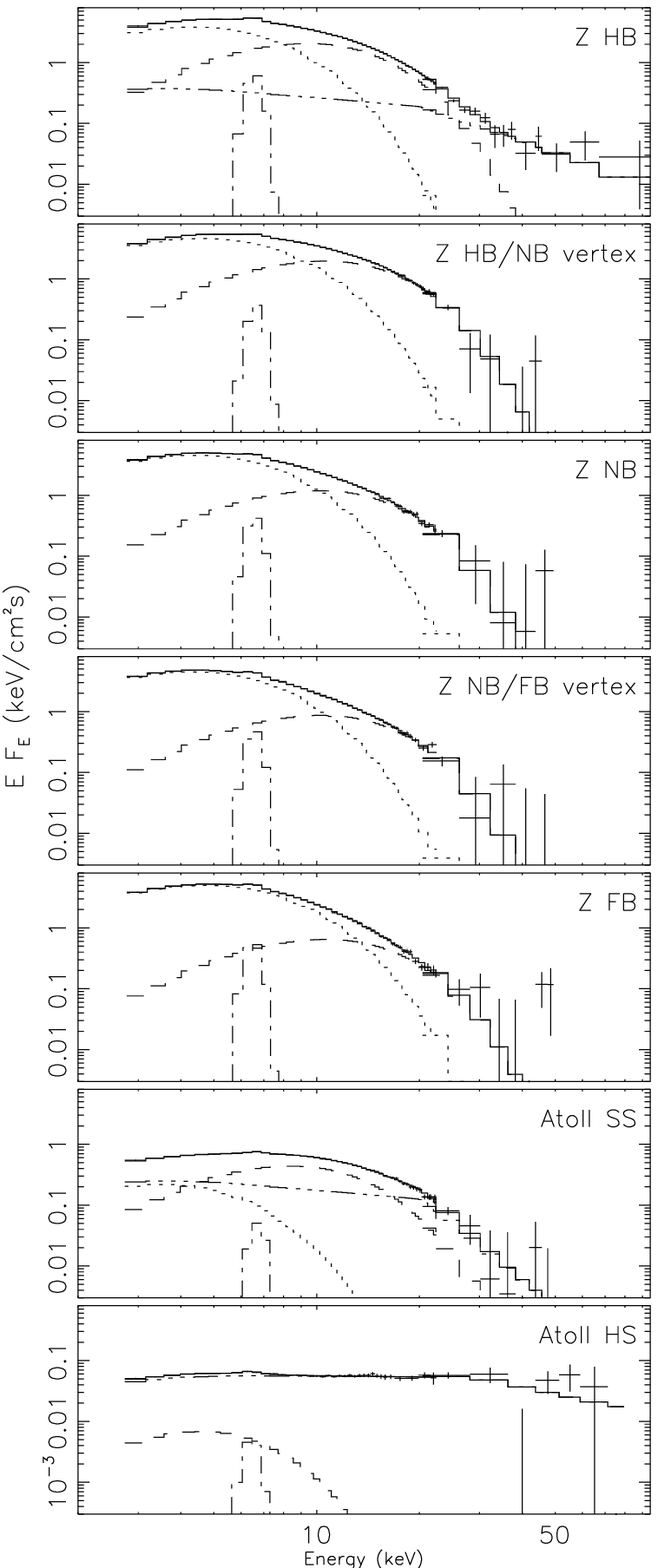} 
\caption{Examples of unfolded spectra at different
states/branches. The total model fit is shown as a solid line, and it
consists of an MCD component (dotted line), a BB (dashed line), a CBPL
component (dot-dot-dashed line), and a Gaussian line (dot-dashed
line). The spectrum at the top panel shows the detection of a hard
tail on the HB.
\label{fig:hardtail}} 
\end{figure}

We used the hybrid model from LRH07 (Model 6) to fit the X-ray spectra
of \object{XTE J1701-462}. In this model, the soft-state spectra are
fit with a combination of MCD, BB, and CBPL. The CBPL component is a
constrained broken power-law ($E_{\rm b} = 20$ keV and $\Gamma_1 \leq
2.5$) that can assume the role of weak Comptonization. The atoll
hard-state spectra are fit with a combination of BB and BPL.

As in LRH07, the PCA and HEXTE pulse-height spectra were fitted
jointly, with the normalization of the HEXTE spectrum relative to the
PCA spectrum allowed to float between 0.7 and 1.3 (the average
best-fitting value is 0.99$\pm$0.12). For the PCA spectra, we used
energy channels 4--50, corresponding to 2.7--23 keV. For the HEXTE
spectra, we used 20.0--80.0 keV for hard-state observations and
20.0--50.0 keV for soft-state ones. A Gaussian line was also included
in the fit, with its central line energy constrained to be between
6.2--7.3 keV (the average best-fitting value is $\sim$6.5 keV) and the
intrinsic width of the Gaussian line ($\sigma$) fixed at $0.3$
keV. Interstellar absorption was taken into account with the hydrogen
column density fixed at $N_{\rm H}=2.0\times 10^{22}$ ${\rm
cm}^{-2}$. The values of the hydrogen column density and the intrinsic
width of the Gaussian line were based on joined fit of simultaneous
observations of this source with {\it Swift} and \xte.

The orbital inclination of the binary system cannot be very high
($\lesssim$75), as eclipses or absorption dips were not observed in
this source. However, the Fe emission line is quite weak (equivalent
width $\lesssim$50 eV), compared to other Z sources
\citep[e.g.,][]{whpeha1986,camira2008}, and this could imply that the
inclination is not very low, either \citep{faiwre2000}. In this paper,
a binary inclination of $70\degr$ was assumed. The luminosity and
radius-related quantities were calculated using a distance of 8.8 kpc,
inferred from type I X-ray bursts that showed photospheric radius
expansion \citep{lihore2007,lihoal2009}.

LRH07 showed that the CBPL component is only required at low-$L_{\rm
X}$ soft-state observations. At higher $L_{\rm X}$, most spectra can
be fit by the MCD+BB model. In
Figure~\ref{fig:xtej1701model5model6com}, we show the fraction of
the $L_{\rm CBPL}$ using steady 960-s spectra in Z stages and all
spectra in the atoll SS (\S\ref{sec:variability}).  The $L_{\rm CBPL}$
was obtained by integrating from 1.5 keV to 200 keV. The choice of the
upper limit is not critical since $\Gamma_2$ is normally ${>}2$. The
majority $({\sim}91\%$) of the steady spectra in the Z stages have
$L_{\rm CBPL}$ that is less than $10\%$ of the total luminosity. There
are two situations when the CBPL component contributes
significantly. The first is when the source intensity is low, at the
end of the Z stage and in the atoll SS. The second is when the source
enters the HB in the Sco-like Z stages or the upturn of the HB in the
Cyg-like Z stage ($\sim$800--1300 counts/s/PCU;
Figure~\ref{fig:sumhid}). In the top panel of
Figure~\ref{fig:hardtail}, we show a sample spectrum with a hard tail
extending above 100 keV, from the upturn of the HB in sample interval
Ia. It is a combination of six observations made on 2006 February 2-3,
with observation IDs 91106-02-02-12 and 91106-02-03-[01--05]. The
combined spectrum has an total exposure of 15 ks, a hard color of 0.39
and an intensity near 1260 counts/s/PCU.

We also fit the steady 960-s spectra in the Z stages using the MCD+BB
model, i.e. without the CBPL to account for a possible hard
tail. About ${90}\%$ of these spectra have an increase in the total
$\chi^2$ smaller than 4.6 (compared to the MCD+BB+CBPL model), a
$90\%$ confidence level criterion for the inclusion of the CBPL in the
final fit for each spectrum. The initial photon index $\Gamma_1$ often
reaches the hard limit in the fit of soft spectra so that the CBPL has
only two free parameters in practice. 

Figure~\ref{fig:hardtail} illustrates the unfolded spectra in
different states/branches. The spectrum at the top panel is the one
that we used to show the detection of a hard tail (see above). The
sample spectra corresponding to the Z-source HB/NB vertex, NB, NB/FB
vertex and FB are all from sample interval IIa, and all have exposures
$\sim$960 s and intensities $\sim$1200 counts/s/PCU. There is no CBPL
component in these unfolded spectra, because the inclusion of such a
component does not improve the $\chi^2$ (see above). The spectrum
corresponding to the atoll-source SS is from observation
93703-01-02-11, with an exposure of 6155 s and an intensity of 181
counts/s/PCU. The spectrum corresponding to the atoll-source HS is a
combination of observations 93703-01-03-14 and 93703-01-03-16, with an
exposure of 13 ks and an intensity of 17 counts/s/PCU.

The spectral fit results are shown in the following sections. For
clarity, we only plot data points with small error bars, i.e., if the
difference between the upper and lower limits ($90\%$ confidence) of
the temperature of the thermal components is smaller than 0.7 keV. In
the end, $1.1\%$ and $2.4\%$ of the data points for the MCD and BB
components are omitted, respectively.

\subsection{Atoll Source Stage}
\label{sec:atollmodelfit}

\begin{figure*}\epsscale{1.0}
\plotone{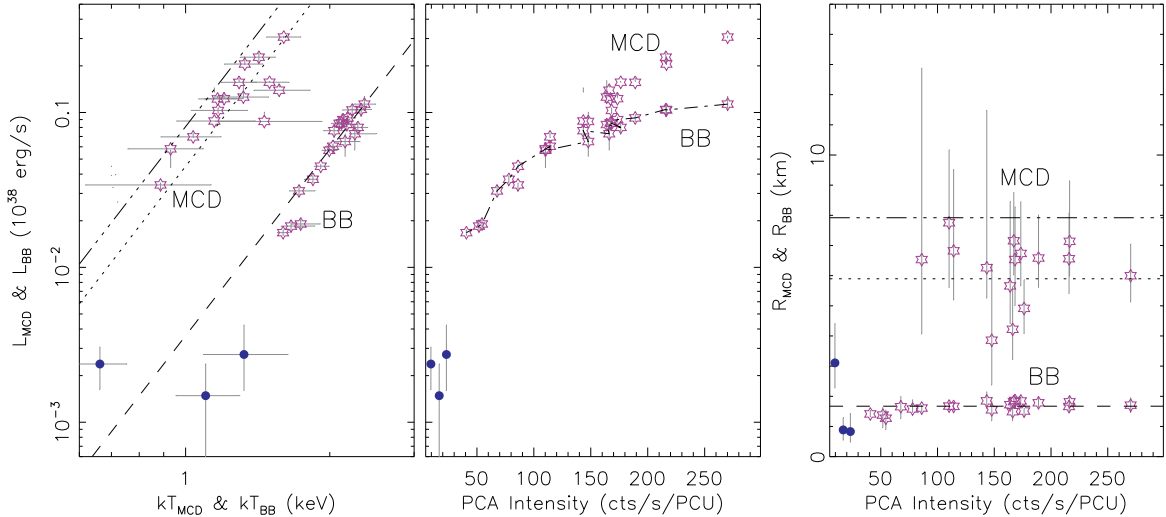}
\caption{Spectral fitting results for
the atoll stage of the outburst (stage V), using MCD+BB+CBPL for the
SS and BB+BPL for the HS. There is no MCD component for the HS, and
the blue filled circles correspond to the BB in all panels. The
luminosities of the thermal components (MCD/BB) are plotted against their
characteristic temperatures and the PCA intensity in the left and
middle panels, respectively. The data points for the BB component in
the SS in the middle panel are connected by a dot-dashed line for
clarity. The right panel shows the characteristic emission size of the
thermal component versus the PCA intensity. The dot-dot-dashed line
corresponds to the size of the NS inferred from type I X-ray
bursts. The dotted and dashed lines correspond to $R=6.0$ km and
$R=1.7$ km, respectively, assuming $L_{\rm X}=4\pi R^2\sigma_{\rm SB}
T^4$.
\label{fig:lum_Tbb_atoll}}
\end{figure*}

Since our spectral model was developed for observations of transient
atoll sources, we begin the fitting process of \object{XTE J1701-462}
with the atoll stage (V). We compare our results with those from
LRH07, to see whether the spectral evolution in stage V is consistent
with the behavior of other atoll sources. The CD and HID for this
stage are shown in the bottom panel of
Figure~\ref{fig:ccdiagfivestages}. The spectral fits show that the
fraction of the $L_{\rm CBPL}$ can reach more than $30\%$ in the SS
(Figure~\ref{fig:xtej1701model5model6com}). The HS observations
are very faint and are dominated by the BPL component ($\sim$$95\%$)
with initial photon index $\sim$2.  The spectral fit results for the
MCD and BB components are shown in Figure~\ref{fig:lum_Tbb_atoll}.

The left panel shows the luminosity of each thermal component versus
its color temperature, $kT_{\rm BB}$ or $kT_{\rm MCD}$. For reference,
we also show the lines for constant radius, assuming $L_{\rm
X}=4\pi\sigma_{\rm SB}R^2T^4$. The NS radius (8 km), inferred from
type I X-ray bursts for an assumed distance of 8.8 kpc
\citep{lihoal2009}, is shown with dot-dot-dashed lines in
Figure~\ref{fig:lum_Tbb_atoll}. The dot-dashed lines correspond to $R
= 6.0$ km, and the dashed lines to $R = 1.7$ km. These latter two
values are derived from the fit to the obtained best-fitting $R_{\rm
MCD}$ and $R_{\rm BB}$ values, respectively. The inner disk radius is
comparable with the inferred NS radius. However, we note that these
values have large systematic uncertainties, since accurate
measurements require knowledge of the distance, inclination, and other
parameters such as the hardening factors for the disk and burst
spectra.

The most remarkable result seen in Figure~\ref{fig:lum_Tbb_atoll} is
that the disk and boundary components evolve roughly along $L_{\rm X}
\propto T^4$ tracks, consistent with LRH07 results for \object{Aql
X-1} and \object{4U 1608-52}. Moreover, the boundary layer is small
and remains nearly constant in size throughout the SS. This can be
seen in the right panel of Figure ~\ref{fig:lum_Tbb_atoll}, which
shows the best-fitting $R_{\rm MCD}$ and $R_{\rm BB}$ versus the
source intensity. $R_{\rm BB}$ cannot be constrained very well in the
HS, but its values are marginally consistent with those from the SS.

In the middle panel of Figure~\ref{fig:lum_Tbb_atoll}, we show the
luminosity of each component (the MCD and BB) versus the PCA
intensity. This plot makes it easy to link the spectral results to the
HID in Figure~\ref{fig:ccdiagfivestages}. To help distinguish the two
spectral components, the data points for the BB results in the SS are
connected by a dot-dashed line.

\subsection{Z Source Stages}
\label{sec:zmodelfit}

\begin{figure*}\epsscale{1.0}
\plotone{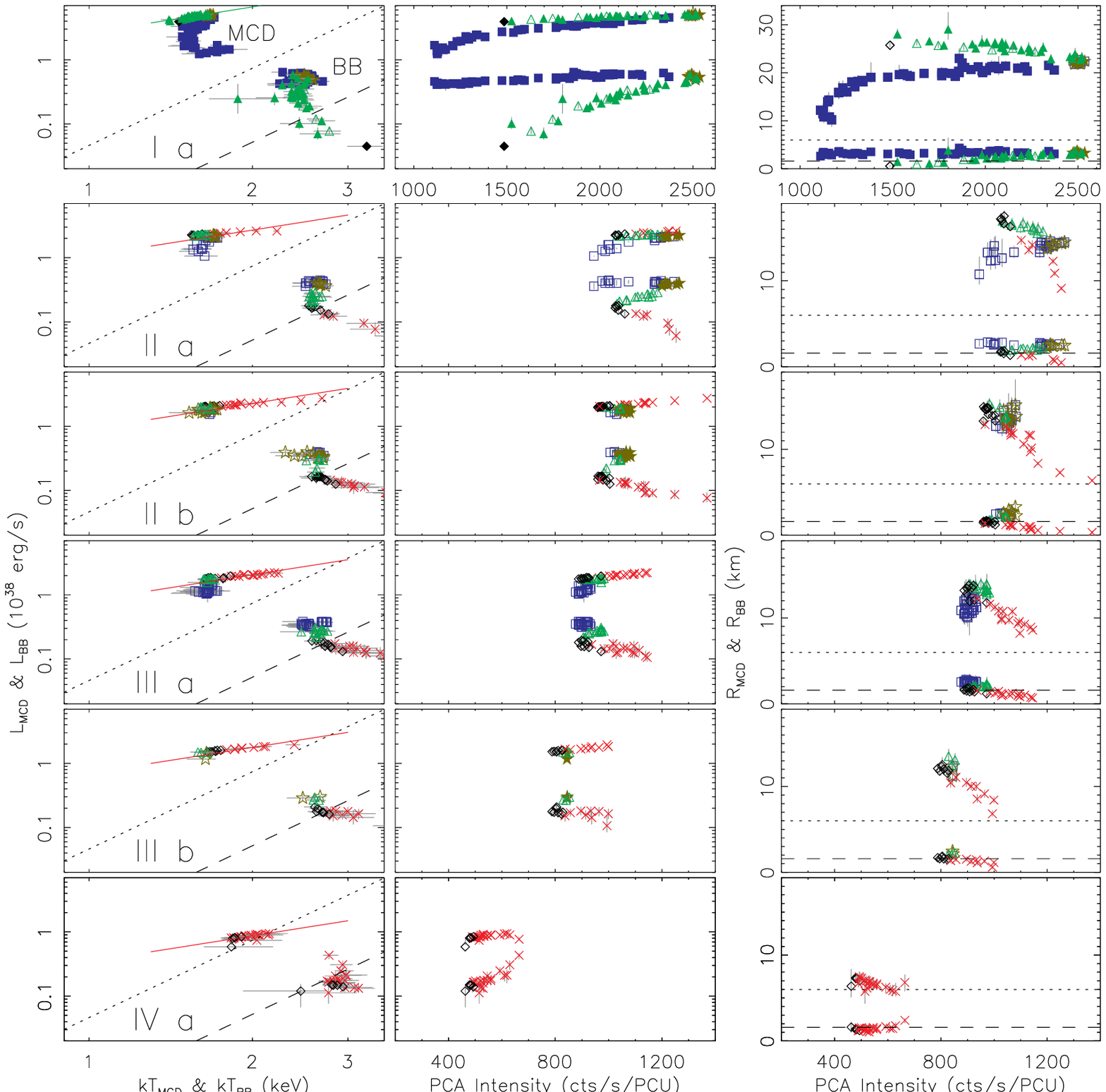}
\caption{The spectral fitting results for the sample intervals (see
\S\ref{sec:sampleintervals1}). As in Figure~\ref{fig:lum_Tbb_atoll},
the luminosities of thermal components (MCD/BB) are shown versus their
characteristic temperatures and versus the PCA intensity. The right
column shows the characteristic emission size of the thermal component
versus the PCA intensity. Each row of panels corresponds to one sample interval.
The dotted line and dashed line in Figure~\ref{fig:lum_Tbb_atoll} are
replotted here for reference. In each panel, results of the MCD are
always at the top, and those of the BB at the bottom.
\label{fig:lum_Tbb_sub}}
\end{figure*}

\begin{figure*}\epsscale{1.0}
\plotone{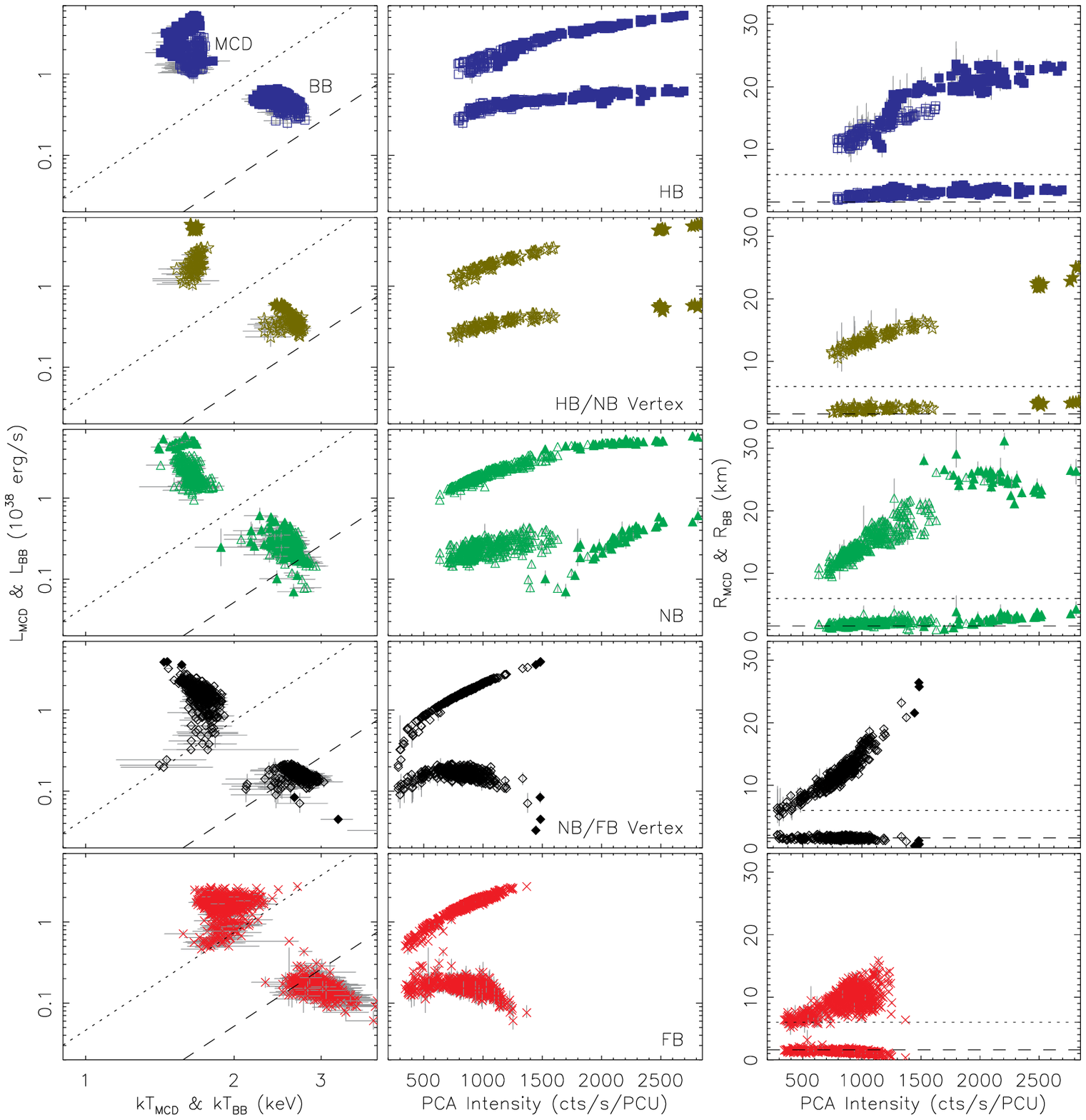} 
\caption{The same as Figure~\ref{fig:lum_Tbb_sub}, but combining all
observations throughout the Z stages of the outburst. From the top to
the bottom, the rows correspond to the HB, HB/NB vertex, NB, NB/FB
vertex, and FB respectively. The results are derived from the steady
960-s spectra plus the spectra created from the boxes shown in
Figure~\ref{fig:colors_sub}.
\label{fig:lum_Tbb_all2}} 
\end{figure*}

The spectral evolution in the Z stages is more complex. To understand
the physical processes that drive the evolution of Z sources along
different branches, we carried out spectral fits for the samples of
the Z tracks that were discussed earlier (time intervals marked in
Figure~\ref{fig:lctwobands} and light curves, CDs and HIDs shown in
Figure~\ref{fig:colors_sub}). As explained in
\S\ref{sec:sampleintervals1}, we fit steady 960-s spectra directly,
and we use box selections in the HIDs to combine 32-s spectra
accumulated from the non-steady intervals. The spectral fitting
results for the sample intervals are shown in
Figure~\ref{fig:lum_Tbb_sub}, one row for each interval. The
quantities plotted in the three panels of each row are the same as for
Figure~\ref{fig:lum_Tbb_atoll}. The results are discussed below, one Z
branch or vertex at a time.

In addition to our consideration of these sample intervals, we also
fit all the steady 960-s spectra of the four Z stages. The results are
shown in Figure~\ref{fig:lum_Tbb_all2} for each
branch/vertex. Some spectral results for non-steady observations in
the sample intervals are also included in this figure. The dotted and
dashed constant radius lines from Figure~\ref{fig:lum_Tbb_atoll} are
again shown in Figures~\ref{fig:lum_Tbb_sub} and
\ref{fig:lum_Tbb_all2} for reference.

\subsubsection{The NB/FB Vertex}
\label{sec:nfb vertex}

\begin{figure}
\plotone{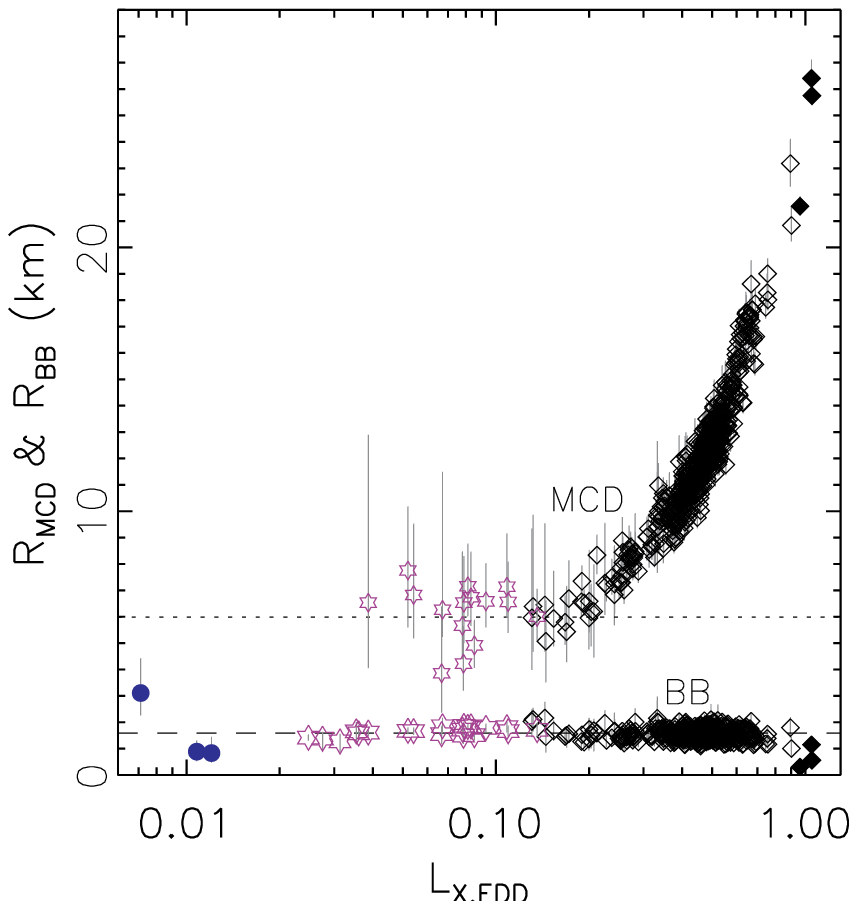} 
\caption{The emission sizes of the thermal components versus the total
$L_{\rm X}$ for the NB/FB vertex and the atoll source stage V.
\label{fig:R_LforV}} 
\end{figure}

We begin the assessment of our spectral results for the Z stages by
considering the evolution of the NB/FB vertex (black diamond symbols),
which is globally shown to be well organized in
Figure~\ref{fig:lum_Tbb_all2}. When the Z-source behavior departs
from the atoll track (Figure~\ref{fig:lum_Tbb_atoll}), the disk no
longer follows the $L_{\rm MCD} \propto T_{\rm MCD}^{4}$ relationship,
and $R_{\rm MCD}$ begins to increase with intensity. 

To see this effect more clearly, we plot the radii of the thermal
components versus the total $L_{\rm X}$ in units of $L_{\rm EDD}$ for
the NB/FB vertex in Figure~\ref{fig:R_LforV}, with the results from
the atoll track in stage V also included for comparison. $L_{\rm EDD}$
is taken to be $3.79\times 10^{38}$ erg/s. This value was estimated to
be the Eddington luminosity for NS type I X-ray bursts showing
photospheric radius expansion \citep{kudein2003}. This value was also
used to infer the distance of 8.8 kpc for \object{XTE J1701-462}
\citep{lihoal2009}. We note that the same results would be
obtained if we take the net flux from the spectral fit, while
correcting the MCD term by inclination effect, and then divide by
the average maximum flux measured for the two radius expansion
bursts. Finally, we note that $L_{\rm X,EDD}$ is a quantity that is
useful for scaling purposes only, since proper considerations of the
Eddington limit should consider the disk and boundary layer of Z
sources separately (\S\ref{sec:discussion}), and, in addition, it is
likely that a portion of the BB flux is obscured by the disk (LRH07).

In Figure~\ref{fig:R_LforV}, we see that the inner disk radius remains
constant, at a value presumed to represent the innermost stable
circular orbit (ISCO), until $L_{\rm X}$ reaches ${\sim}0.2$ $L_{\rm
EDD}$. Above this luminosity, the inner disk radius increases with
luminosity. This is quite possibly a signature of the local Eddington
limit in the disk and will be further discussed in
\S\ref{sec:discussion}. When \object{XTE J1701-462} deviates from the
$L_{\rm MCD} \propto T_{\rm MCD}^{4}$ track, its evolution is
consistent with $L_{\rm MCD} \propto T_{\rm MCD}^{-4}$, or
equivalently $L_{\rm MCD} \propto R_{\rm MCD}$, with $T_{\rm MCD}$
slightly decreasing with $L_{\rm MCD}$.

In contrast to the behavior of the disk radius, the boundary layer
($R_{\rm BB}\sim 1.7$ km) maintained its small (nearly constant) size
from the atoll stage to all observations in the NB/FB vertex, with
luminosity ranging from $\sim$0.01 to 1 $L_{\rm EDD}$. These results
suggest an intimate relation between the atoll track and the NB/FB
vertex.

\subsubsection{The FB}
\label{sec:FB}
The behavior of the FB is evaluated from the results of the sample
intervals, shown in red cross symbols in
Figures~\ref{fig:lum_Tbb_sub}. Besides the two constant radius lines,
we also plot solid lines for which $L_{\rm MCD} \propto T_{\rm
MCD}^{4/3}$. These lines describe the relationship between disk
luminosity and inner disk temperature when the source has a variable
inner disk radius at a constant accretion rate, as shown in the
following. The disk temperature $T$ at radius $R$ is \citep{ha1981}
\begin{equation}
T(R)=\left(\frac{3GM\dot{m}}{8\pi \sigma_{\rm SB} R^3}\right)^{1/4}
\label{eq:MCDTR}
\end{equation}
where $M$ is the mass of the NS. Evaluating the above equation at the
inner disk radius and considering that $L_{\rm MCD}=4\pi\sigma_{\rm
SB}R_{\rm MCD}^2 T_{\rm MCD}^{4}$ \citep{miinko1984}, we obtain
\begin{equation}
L_{\rm MCD}=4\pi\sigma_{\rm SB}\left(\frac{3GM\dot{m}}{8\pi \sigma_{\rm SB}}\right)^{2/3}T_{\rm MCD}^{4/3}.
\label{eq:constantmdot}
\end{equation}
A constant $\dot{m}$ then leads to $L_{\rm MCD} \propto T_{\rm MCD}^{4/3}$, as represented by the red solid lines in Figure~\ref{fig:lum_Tbb_sub}.

This figure shows that the FB tracks closely follow the red solid
lines, implying that the disk evolution is consistent with an inner
disk radius varying under the condition of a constant $\dot{m}$. In
the NB/FB vertex the disk is truncated at a larger radius than in the
atoll stage. The FB is traced out when the disk refills temporarily
and the inner radius shrinks to the value seen in the atoll stage,
which is presumed to be the ISCO. As $\dot{m}$ increases, the disk in
the NB/FB vertex is truncated at a progressively larger radius,
thereby shifting and lengthening the constant $\dot{m}$ line along
which the disk evolves on the FB.  The superposition of these tracks
can be seen in the left bottom panel of
Figure~\ref{fig:lum_Tbb_all2}.

The BB component contributes less as the source ascends the FB,
especially in the intervals from stages II--III. At the same time, the
temperature appears to increase, while the effective radius sharply
decreases. We have no interpretation for these effects, and it is also
possible that the results are affected by systematic problems with our
model when one component (BB, in this case) contributes a small
fraction of the flux. The FB of interval Ia is of the dipping type,
and its spectra cannot be fit well by the MCD+BB+CBPL
model. Therefore, the results for the FB of this interval are not
shown.

\subsubsection{The NB}
\label{sec:NB}

\begin{figure}
\plotone{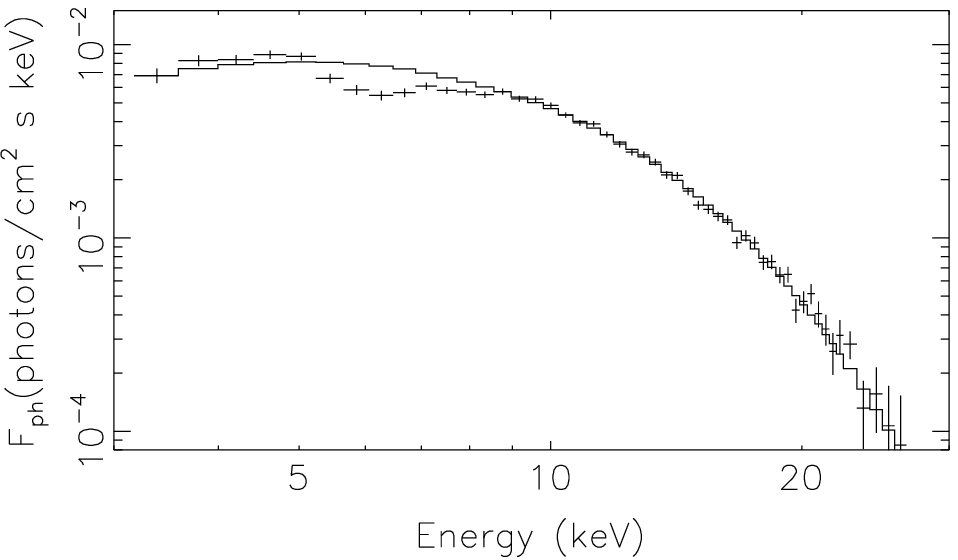} 
\caption{The difference of spectra for the two ends of the NB during
interval IIIa. The spectrum is fit with a BB model (solid line)
excluding the region 4.5--8 keV around the Fe line. The success of
this model confirms that the evolution along the NB during this
Sco-like Z stage is mostly due to the changes of the boundary layer
emission area.
\label{fig:NHsubpha}} 
\end{figure}

Results from the sample intervals (Figure~\ref{fig:lum_Tbb_sub}) show
that, as the source ascends the NB in the Sco-like stages II--III, the
MCD component seems to remain unchanged, while the BB radius increases
at constant temperature. The latter effect causes an increase in the
source intensity at energies around 10--30 keV, with almost no change
at energies $<$7 keV (green dashed lines in
Figure~\ref{fig:twopha_pcahexx1701}). Since the NB is mainly the
result of a changing BB radius, it is interesting to see whether the
difference of the spectra from the two ends of the NB can be fit by a
BB model. We test this on the NB in sample interval IIIa. Excluding
the region around the Fe line near 6.4 keV, the resulting spectrum
(Figure~\ref{fig:NHsubpha}) can indeed be fit by a BB, with
temperature $kT_{\rm BB}=2.71\pm 0.03$ keV. This temperature is close
to that of the individual spectra on this particular NB, confirming
that the source evolution along the NB is mostly due to an increase in
the normalization of the BB at constant temperature. The difference
spectrum shows a dip around the Fe line, indicating that the Fe line
emission is stronger at the lower part of the NB, despite the fact
that the source intensity at the lower part of the NB is lower.

On the NB of interval Ia the BB still changes very significantly, but
in that case there is also a clear change in the MCD parameters,
roughly consistent with evolution at a constant $\dot{m}$ through the
disk (Figure~\ref{fig:lum_Tbb_sub}). From the NB/FB vertex toward the
HB/NB vertex, $L_{\rm MCD}$ and $kT_{\rm MCD}$ increase while $R_{\rm
MCD}$ decreases. The $L_{\rm BB}$ and $R_{\rm BB}$ also increase. All
this results in a large increase in the source intensity, at energies
around 10-30 keV (Figure~\ref{fig:twopha_pcahexx1701}).

The possible origins of the BB increase on the Sco-like NB are
discussed in \S\ref{sec:discussion}. The coupled changes in the MCD
and BB components for the Cyg-like NB seem to be complicated, and we
have no simple picture of the corresponding physical changes.

\subsubsection{The HB/NB Vertex}
\label{sec:HNB vertex}

\begin{figure}
\plotone{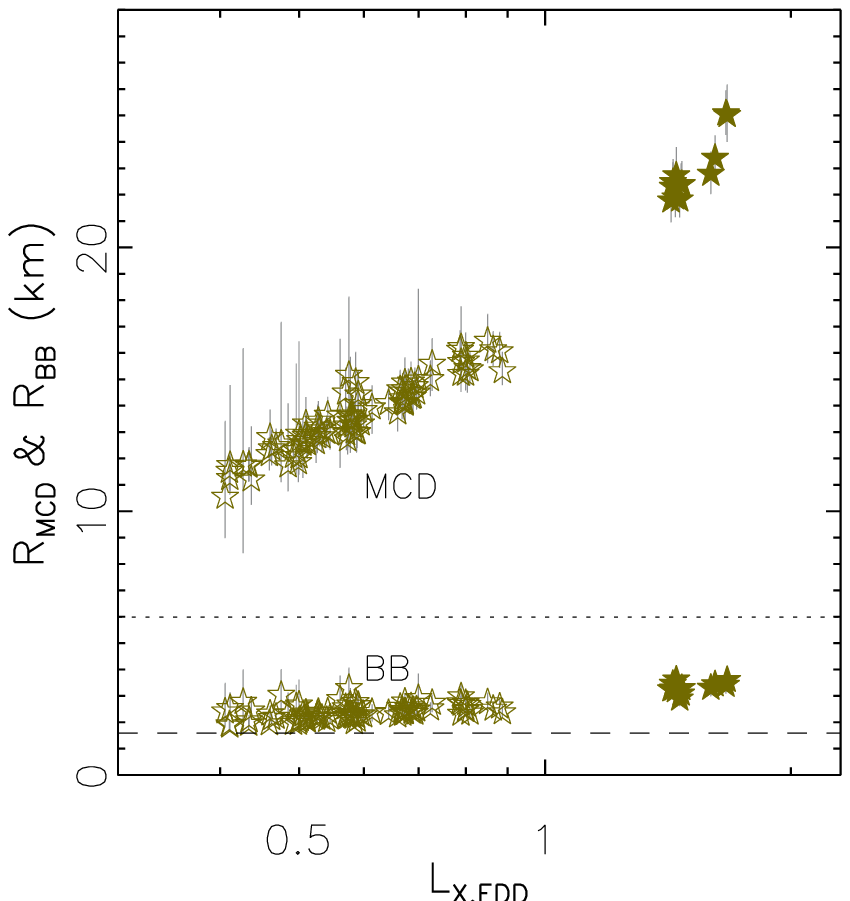} 
\caption{The emission sizes of the thermal components versus the total
$L_{\rm X}$ for the HB/NB vertex.
\label{fig:R_LforU}} 
\end{figure}

Similar to the NB/FB vertex, the behavior of HB/NB vertex in the HID
is well organized, i.e., it evolves along a single line in
Figure~\ref{fig:sumhid}. Since there is little change in the MCD
component on the NB of the Sco-like stages, the differences between
the MCD quantities of the HB/NB vertex and those of the NB/FB vertex
are small (Figure~\ref{fig:lum_Tbb_sub}). Only when the luminosity
increases above a certain value do their differences become
significant. This is reflected in the global spectral results for the
HB/NB vertex in Figure~\ref{fig:lum_Tbb_all2}: $kT_{\rm MCD}$ in
the HB/NB vertex hovers around 1.6 keV, whereas $kT_{\rm MCD}$ in the
NB/FB vertex shows a slight decrease at high luminosity.

Both $R_{\rm BB}$ and $R_{\rm MCD}$ increase with intensity. The
visible BB effective radius always appears smaller than the disk
radius and never reaches the values of the inner disk radius seen in
the atoll stage. Both $L_{\rm MCD}$ and $L_{\rm BB}$ increase with
intensity, but $L_{\rm MCD}$ increases much faster.

Although the Cyg-like and Sco-like Z tracks are quite different, the
data points of HB/NB vertex in Cyg-like Z stage I seem to lie along
the high-$L_{\rm X}$ extension of the points of the HB/NB vertex in
the Sco-like Z stages II--IV. Figure~\ref{fig:R_LforU} shows the
emission sizes $R_{\rm MCD}$ and $R_{\rm BB}$ of the thermal
components with respect to the total $L_{\rm X}$, instead of the
intensity. The HB/NB vertex extends from $\sim$0.4 to 1.7 Eddington
luminosity, and the $R_{\rm MCD}$ and $R_{\rm BB}$ both increase with
luminosity. However, there is a gap between 0.9--1.4 $L_{\rm EDD}$.

\subsubsection{The HB}
\label{sec:HB}

\begin{figure}
\plotone{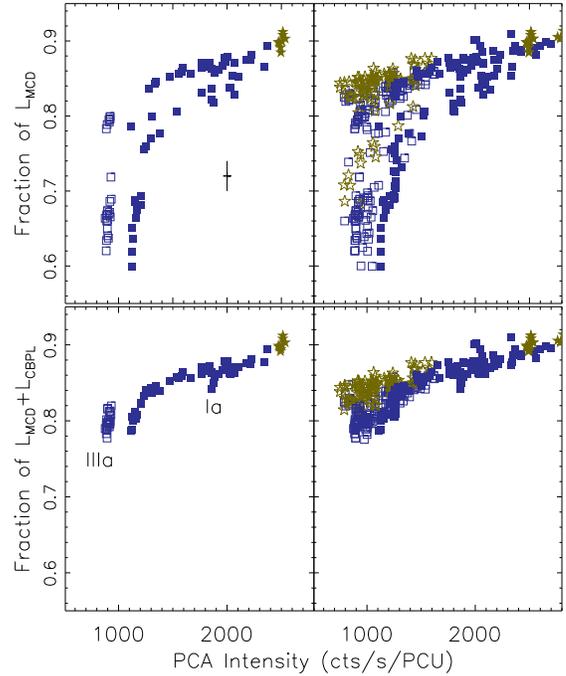} 
\caption{The fractions of the $L_{\rm MCD}$ (upper panels) and $L_{\rm
MCD+CBPL}$ (lower panels) on the HB and the HB/NB vertex. Left panels
show sample intervals Ia and IIIa, while the right panels show the entire
outburst.
\label{fig:HBmcdfraction}} 
\end{figure}

The HB is first investigated using the sample observations
(Figure~\ref{fig:lum_Tbb_sub}). In general, $L_{\rm MCD}$ decreases as
the source ascends the HB away from the HB/NB vertex. This causes
changes in the spectrum at low energies, as shown in
Figure~\ref{fig:twopha_pcahexx1701} (blue dotted lines). The BB
component, however, varies little. In contrast with the decrease in
the intensity at low energy, there is an increase in the intensity at
high energies, going out along the HB. This is mostly due to an
increase of the Comptonized component
(Figure~\ref{fig:xtej1701model5model6com}), which for the Cyg-like
Z stage mostly occurs on the upturn of the HB. The significant detection
of Comptonization on the HB of the Z stages is consistent with a
series of discoveries of hard tails in other Z sources
\citep{pafati2006,fafrza2005,diroia2001,dahero2001,distro2000},
although there are claims of hard tail detections on branches other
than the HB in some of these reports.

To illustrate the coupled behavior between the MCD and the CBPL
components on the HB more quantitatively, we plot in
Figure~\ref{fig:HBmcdfraction} the fractions of the $L_{\rm MCD}$
(upper panels) and the $L_{\rm MCD}+L_{\rm CBPL}$ (lower panels) for
sample intervals Ia and IIIa. The results for all observations on the
HB and HB/NB vertex are also plotted, on the right. We can see that
while the fraction of the $L_{\rm MCD}$ on the HB changes by $20\%$
(upper panels), the fraction of the $L_{\rm MCD}+L_{\rm CBPL}$
maintains a much smoother track over the Sco-like Z tracks. On shorter
timescales, the quantity $L_{\rm MCD}+L_{\rm CBPL}$ changes also very
little over a typical Sco-like HB, e.g., only 5$\%$ for interval
IIIa. The above results imply that in the Sco-like Z stages, as the
source climbs up the HB, thermal emission from the disk is converted
into a hard component by Comptonization. This is also true for the
upturn of the HB in the Cyg-like Z stage. Thus the similarity between
the Sco-like HB and the upturn of the Cyg-like HB is supported by both
the fraction of Comptonization and the coupling between the MCD and
CBPL luminosities.

\section{Broadband Variability}
\label{sec:timing}

\begin{figure}
\plotone{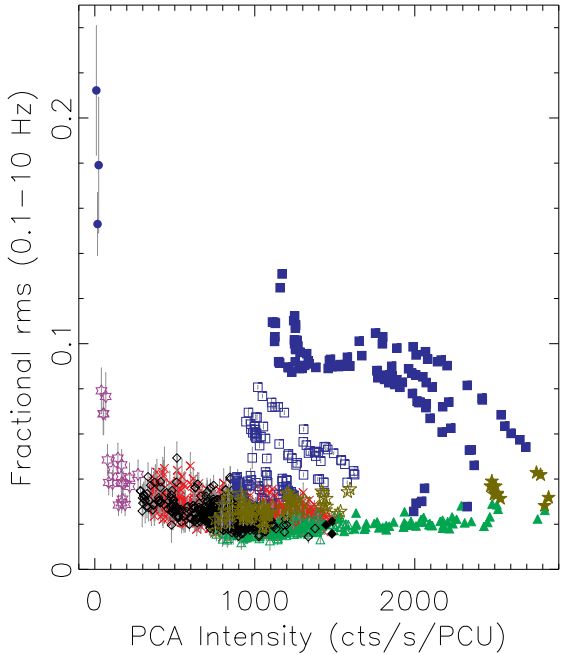} 
\caption{The rms from 0.1--10 Hz for the entire outburst. Elevated
continuum power (i.e., rms $>$ 5$\%$) in the PDS is limited to the Z
source HB and atoll source HS.
\label{fig:rmsall}} 
\end{figure}

Although this work mostly focuses on the spectral properties of
\object{XTE J1701-462}, timing properties are important tools for
understanding the evolution along atoll and Z tracks. In LRH07 we used
a measure of the broadband variability to compare the results of our
spectral fits to those obtained for black hole systems. In particular,
we focused on the relation between the fraction of Comptonized flux
and the strength of the variability, as measured by the integrated
(fractional) rms in the power density spectrum (PDS; 0.1--10 Hz).

In Figure~\ref{fig:rmsall} we plot the integrated rms versus the PCA
intensity for \js.  Similar to the classical atoll sources, the power
is very weak in the atoll SS (${<}6\%$) and increases in the atoll
HS. In the Z stages, except for the HB, the power is always very weak,
with typical rms of $2\%$. Although the Sco-like HB typically has a
small intensity range ($\sim 100$ counts/s/PCU), the rms can increase
quite significantly, by up to $4\%$, from the HB/NB vertex to the tip
of the HB. We attribute this rms increase to the growing importance of
Comptonization on the HB. This is consistent with the result in LRH07,
in which we found that the rms is strongly tied to the fraction of the
Comptonized component for both atoll sources and black hole binaries.

On the upturn of the Cyg-like HB the rms increases significantly,
another aspect that makes it similar to the Sco-like HB. However, the
rms on the upturn of the Cyg-like HB is higher than on the Sco-like
HB. Moreover, on the non-upturn part of the Cyg-like HB, the rms
already increases strongly, even though the Comptonization fraction
remains low (Figure~\ref{fig:HBmcdfraction}). The analysis of the
frequency resolved spectroscopy on the typical Cyg-like Z sources
\citep{giremo2003} indicated that this boost in rms seems to arise
from the boundary layer.

\section{DISCUSSION}
\label{sec:discussion}
\subsection{Secular Evolution of XTE~J1701-462 and the Role of $\dot{m}$}
\label{sec:secchange}
During its 2006-2007 outburst, \js\ successively shows characteristics
of Cyg-like Z, Sco-like Z, and atoll sources, and the stages for each
source type are clearly correlated with the X-ray luminosity. The Z
tracks move substantially in the HID, while creating distinct lines
for the upper (HB/NB) vertex and the lower (NB/FB) vertex,
respectively (Figure~\ref{fig:sumhid}). The line for the lower vertex
smoothly connects to the atoll track. \js\ shows a large dynamic range
in luminosity that is typical of X-ray transients, and the overall
shape of the broad band light curve is expected to represent the
temporal evolution of $\dot{m}$. It is then straightforward to examine
Figure~\ref{fig:asmdata} and Table~\ref{tbl-1} and conclude that the
transitions between source types and the secular changes in the HID
are both driven by the changes in $\dot{m}$.

A more detailed accounting of $\dot{m}$ can be derived from the
results of our particular X-ray spectral model, which tracks the
contributions from the MCD, the BB, and weak Comptonization. We first
focus on the MCD, since the disk radiation is always the strongest
component in the spectrum, except for the atoll hard state, and the
origin of the radiation is upstream of any effects of the Eddington limit.
We can infer $\dot{m}$ from the MCD model through \citep{frkira1985}
\begin{equation}
\dot{m}=\frac{2L_{\rm MCD}R_{\rm MCD}}{GM}.
\label{eq:mdot}
\end{equation}
Figure~\ref{fig:lum_Tbb_all2} shows that for each of the vertices,
both $L_{\rm MCD}$ and $R_{\rm MCD}$ increase with intensity, implying
that $\dot{m}$ changes monotonically along each of the two vertex
lines. If we use $\dot{m}$ for the MCD at the upper luminosity end of
the atoll stage as a reference value, then there is a factor $\sim$30
increase in $\dot{m}$ in the lower vertex at the upper luminosity end
of the Sco-like Z stage (i.e., near $1400$ counts/s/PCU), and factors
$\sim$40 and $\sim$60 increase at the brightest points in the lower
and upper vertices, respectively, of the Cyg-like Z stage. These ratios
are larger than the corresponding ratios in luminosity or PCA
intensity, because, for the same $\dot{m}$, an increasingly truncated
disk produces radiation with decreasing efficiency.  These accretion
rates would amplify the sense in which \js\ is seen to be an
extraordinary X-ray transient.

Our spectral results for the MCD also provide insights as to how the
Eddington limit might affect the behavior of Z source. In the atoll
stage, $R_{\rm MCD}$ remains constant, and the value is presumably
close to the radius of the ISCO. In the Z stages, $R_{\rm MCD}$
increases with $L_{\rm MCD}$ (and $\dot{m}$) along the NB/FB vertex
line (Figure~\ref{fig:R_LforV}). One may interpret this result as the
effect of the local Eddington limit in the disk. The Eddington limit
is reached when the radiation pressure overcomes the force of gravity.
For a spherical case, both gravity and radiation forces vary as
$r^{-2}$, and the Eddington limit has a single value that covers all
radii. However, in a disk system, both the gravity and radiation
forces depend more complicatedly on $r$, and the Eddington limit
should be reached locally \citep{ka1980,fu2004}. In a standard disk
(Equation~\ref{eq:MCDTR}), the locally generated luminosity varies as
$r^{-3}$, and radiation pressure most effectively moves matter in the
vertical direction, while the vertical component of gravity force
roughly varies as $r^{-2}$, if we assume that the thickness of the
disk scales linearly with $r$. When $\dot{m}$ increases to a certain
point, the inner disk must adjust to the radius where the local
Eddington limit is reached, while at larger radii the disk can
continue to produce thermal radiation \citep{ka1980,fu2004}.

Detailed considerations of super-Eddington mass flows in accretion
disks have shown $\dot{m}$ regimes in which the disk is thickened, an
advective or quasi-radial accretion flow dominates the region inside
the inner disk radius, and substantial mass may be driven out of the
system \citep{ka1980,wafuta2000,mikata2000,fu2004,ohmi2007}. Such
accretion solutions are often referred to as ``slim disk'' models.
Structural changes for the slim disk will modify the $T(r)$ function,
so that the emergent spectrum will no longer resemble the MCD model.
However, the divergence between these models might not be apparent
until $\dot{m}$ is much larger than the critical value that first
brings the inner disk to the Eddington limit at the ISCO
\citep{mikata2000}.

Slim disk models depend on the mass of the compact object and the
accretion rate, and there are additional considerations required for
effects of general relativity and of radiative transfer through the
thick disk. Furthermore, such models are usually applied to accreting
black holes, which are free from the additional emission from the
boundary layer and its illumination of the inner disk. In black hole
studies, observers look for evidence of slim disks in high-luminosity
soft-state observations in which the MCD model does not fit the data
well, while a more generic disk model based on the function, $T
\propto r^{-p}$, constrains $p$ to be somewhat lower than the value
($p = 0.75$) required for the MCD \citep{okebka2006}. Application of a
slim disk model for \js, which requires careful considerations of the
NS boundary layer, is beyond the scope of this investigation. We note,
however, that we do not see a $\chi^2$ barrier for our spectral model
when the source is very bright, which suggests that the deviations
between the disk spectrum and the MCD model are not large in the
observed bandpass (3--50 keV).

We have seen that both the upper and the lower vertices evolve along
two distinct lines respectively in the HID over a large overlapping
intensity range (Figure~\ref{fig:sumhid}). However, the variations of
the MCD and BB components along the upper vertex line are different
from those along the lower vertex line
(Figure~\ref{fig:lum_Tbb_all2}). This might imply that the system
is able to respond to the variations in accretion rates in two
different ways along these two vertex lines. We have discussed two
possible disk solutions above (i.e., standard thin disk versus slim
disk), and thus it is possible that the two vertices assume these two
different solutions for disk accretion respectively. This idea will be
further discussed in the next section in terms of the behavior of the
source along the NB, which bridges these two vertices.

Our model fits additionally track the apparent conditions in the NS
boundary layer, via the spectral parameters for the BB component. At
the lower vertex, it is apparent that $L_{\rm BB}$ increases much more
slowly than $L_{\rm MCD}$ and even decreases at the highest intensity
in the Z-source stages (Figure~\ref{fig:lum_Tbb_all2}). In
addition, there is no evidence for BB radius expansion during the
evolution that spans the atoll stage and the lower vertex in the Z
stages (Figure~\ref{fig:R_LforV}). Thus, we can find no clear evidence
on the lower vertex track for any critical point at which Eddington
limit is reached in the BB, which is expected to expand and evolve in
spectral shape as the luminosity passes through the Eddington limit
\citep{insu1999,posu2001}. This casts doubt on the idea that the
gradual expansion of the disk with increasing luminosity at the lower
vertex might be some type of disruption caused by the emission from
boundary layer. On face value, our spectral results imply that
significant mass outflow from the inner disk edge (in the Z stages)
limits $L_{\rm BB}$ to levels below or near the Eddington limit at the
surface of the NS.

Other investigations have looked for an Eddington signature in the
behavior of the BB temperature. A value of $kT_{\rm BB} \sim 2.4$ keV
was found to be a high-luminosity limit for accreting NSs that were
studied with the Fourier-frequency resolved spectroscopy technique
\citep{giremo2003,regi2006}.  This value is close to the peak
temperature seen in radius expansion bursts \citep{gamuha2006}, while
a peak value $\sim$2.7 keV is found for two such bursts from \js\
\citep{lihoal2009}. In both Z-vertices, the NB, and the HB
(Figure~\ref{fig:lum_Tbb_all2}), $kT_{\rm BB}$ hovers near 2.7
keV, with a trend toward slightly lower temperature and slightly
larger $R_{\rm BB}$ at highest $L_{\rm BB}$. Thus, unless there is
variable obscuration of the BB region that masks more substantial
changes in conditions there, we surmise that at the lower vertex the
BB hovers near its Eddington limit and that the BB inherits an
increasingly smaller fraction of the $\dot{m}$ that flows through the
disk, as $\dot{m}$ increases.

Once the inner disk radius is set by the Eddington limit in the lower
(NB/FB) vertex, the only branch that tries to reverse this condition
is the FB. In the upper vertex, there is a second evolution track for
secular variations in which both $R_{\rm MCD}$ and $R_{\rm BB}$ expand
with increases in luminosity.  It is therefore possible that the
mechanism for the NB (i.e., the addition of quasi-radial flow or some
other mechanism) might bring the system to super-Eddington conditions
in both the disk and the boundary layer in the upper (HB/NB) vertex,
while only the disk experiences such conditions in the lower vertex.
Our use of the term "super-Eddington" means, respectively, disk $\dot{m}$
in excess of the Eddington limit when the inner disk radius is at the
ISCO, and boundary layer $\dot{m}$ in excess of the accretion limit at
which the optical surface still coincides with the NS surface.

\subsection{Physical Processes Along the Z Branches of XTE~J1701-462}

We have found that spectral evolution along the Sco-like FB is
consistent with a shrinking of the inner disk radius, while $\dot{m}$
remains constant (Figure~\ref{fig:lum_Tbb_sub}). When we combine the
interpretations for the FB and the NB/FB (lower) vertex, with further
consideration of the increasing evolution speed along the FB
(Figure~\ref{fig:colorspeedFB}), we arrive at the following scenario.
For a given value of $\dot{m}$, the Sco-like Z sources accrete matter
through a truncated disk, with inner radius set by a local Eddington
limit (not by the ISCO). The FB is an instability that springs from
the NB/FB vertex, and it represents a temporary push by the inner disk
toward the ISCO. The entire FB can be traced out in $\sim$10 minutes,
but the evolution is much faster at the top than at the bottom of the
FB (Figure~\ref{fig:colorspeedFB}). Variations at timescales of
$\sim$seconds can easily be seen in the light curve at the top of the
FB. Such timescales are of order the viscous timescales of the inner
disk for a viscosity parameter of 0.01 \citep[][Equation
5.69]{frkira1985}. For the dipping FB in the Cyg-like Z stage, limited
data and poor spectral fits prevent us from deriving any conclusions
about its nature.

The NB bridges the upper and lower vertices of the Z. The Sco-like NB
is traced out as the result of changes in the BB emission size, while
the MCD properties remain nearly constant. Similar BB changes are seen
on the Cyg-like NB, but here the disk also changes, shrinking in
radius at constant $\dot{m}$ similar to the changes seen for the
Sco-like FB. Taken literally, an increase of $\dot{m}$ onto the
boundary layer, with no observable change in $\dot{m}$ for the disk,
could suggest the onset of a radial or advective flow as a secondary
accretion component. This would require that the total $\dot{m}$
increases slightly (i.e., by 10\% or less) as the source ascends the
NB from the lower vertex. This conjecture might be of further interest
toward understanding why NS radio jets begin to be seen when a Z
source begins to ascend the NB, while the jets become stronger and
more steady on the HB \citep[][and references below]{mife2006}.  An
alternative explanation for the Sco-like NB is that $\dot{m}$ remains
constant, but the measured BB area increases as the result of the
geometric changes associated with the boundary layer and/or our line
of sight to it. While there are probably other explanations as well,
it is important to point out one important constraint: the range of
the NB at a specific $\dot{m}$ is not arbitrary, as we can see from
the two vertex lines (Figure~\ref{fig:sumhid}). Moreover, the process
responsible for tracing out the NB appears to be another type of
instability, as the source evolves faster on the NB than in the two
vertices (\S\ref{sec:speed}).

In \S\ref{sec:secchange} we pointed out that the upper and lower
vertices might assume two different solutions for the disk accretion,
as they evolve along two distinct lines in the HID over a large
overlapping intensity range, while the best-fitting results for the
two main components MCD and BB are different. In addition, the two
vertices appear to be stable compared with the evolution along the NB,
and the upper vertex is most often seen at higher luminosity
(Figures~\ref{fig:R_LforV} and \ref{fig:R_LforU}). Both the standard
and slim disks are possible stable accretion disk solutions at high
accretion rates with the slim disk associated with higher accretion
rates and an increased amount of radial advection flow (see references
in \S\ref{sec:secchange}). Also considering that the increase of the
BB emission size along the NB might be explained by an additional
radial flow, we hypothesize that in the upper vertex the disk assumes
a slim disk solution while the disk in the lower vertex is a standard
thin disk. In that case, the NB is formed as the additional radial
flow turns on and the source begins the transition to the slim disk.

As demonstrated in \S\ref{sec:HB}, the Sco-like HB is traced out as
the thermal emission in the disk is converted into a hard component by
Comptonization, while the combined luminosity from the disk and
Comptonization (i.e., $L_{\rm MCD}+L_{\rm CBPL}$) remains roughly
constant. In the most variable case, which is sample IIa, the combined
luminosity varies by 10\% relative to the mean HB value, and we have
shown that this early Sco-like HB has some lingering characteristics
of the Cyg-like stage.  Thus it remains quite possible that the
Sco-like HB is traced out at constant $\dot{m}$ through the disk. The
Cyg-like HB consists of a long horizontal line in the HID, plus an
upturn at the far end.  The upturn portion exhibits an increase in
Comptonization, in common with the Sco-like HB.  However, the
non-upturn part of the Cyg-like HB remains puzzling. Its track in the
HID resembles the secular drift in the upper vertex, which might
suggest that it is not a true HB track.  However the power continuum
shows strongly elevated continuum power, which is a clear signature of
the HB, but from that standpoint the lack of Comptonization is very
surprising.  Further investigations of Cyg-like HB tracks for other
sources are needed.  Finally, as noted above, radio emission has been
detected on the HB, presumably to be due to jet formation
\citep{fedaho2007,mimife2007,fehe2000,pelezi1988}. For the Sco-like HB
and the upturn portion of the Cyg-like HB, the association of
increased Comptonization with radio flux follows a general convention
for X-ray binaries. However, while the spectrum in the atoll HS is
dominated by Comptonization, only modest Comptonization fractions are
found for the HB in Z sources.

Given the variations in $R_{\rm MCD}$ along the FB and the
Comptonization on the HB, we must use the value of ($R_{\rm MCD}
\times (L_{\rm MCD} + L_{\rm CBPL})$) to trace $\dot{m}$ through the
disk for Z sources. Our overall conclusion is that for Sco-like Z
tracks, $\dot{m}$ remains constant to the level of 10\% or less
\citep[see also][]{hovajo2002}. The role of $\dot{m}$ in the evolution
along the Z tracks has been in debate for decades
(\S\ref{sec:intro}). In our analyses, what distinguishes the Z
branches is not $\dot{m}$ but the different mechanisms that spring
from the Z vertices, which are the more stable reference points along
the Z track. The three branches are associated with different forms of
spectral evolution, and our physical interpretations are different
from the concept that Z track evolution is driven by changes in any
single parameter.

Unstable nuclear burning was invoked as a mechanism for the FB by
\citet{chbaja2008}, using observations of Cyg-like Z sources. In their
scenario, there are no changes in $\dot{m}$. We cannot determine
whether this explanation works for the Cyg-like dipping FB of
\object{XTE J1701-462}, but it clearly does not work for the FB in the
Sco-like stages. We note that one type I X-ray burst has been observed
in a FB during its decay back to the NB/FB vertex
\citep{lihoal2009}. We further test this explanation on stage IV as
follows. For a solar abundance, the ratio of the nuclear and
gravitational energies is only about 2.5$\%$
\citep{stbi2006}. However, the Sco-like FB is frequent
(Table~\ref{tbl-1}), and the luminosity variations are very strong.
We calculate the net flaring fluence on the FB in stage IV by
subtracting the associated lower vertex flux from each FB flux
measurement and integrating over time, and the non-flaring fluence is
simply the total fluence minus the flaring fluence. Their ratio turns
out to be very high, $\sim$33$\%$ for stage IV.

\subsection{Comparison with Other NS LMXBs} 
Stage I was shown to be very similar to
the typical Cyg-like Z sources, especially \object{GX 5-1} and
\object{GX 340+0}, by \citet{hovawi2007} based on the CDs/HIDs and
variability. The major difference is in the FB. \object{Cyg X-2} is
similar to \object{XTE J1701-462} in that it also has substantial
secular changes and sometimes even changes between source types
\citep{kuvava1996,wivaku1997}. A more detailed comparison between
these two sources could be valuable.

In Sco-like stages II--III, \object{XTE J1701-462} is very similar to
the Sco-like Z sources. The color and spectral similarities between
\object{XTE J1701-462} and the typical Sco-like Z sources, shown in
Figures~\ref{fig:twopha_pcahexgx340gx17} and
\ref{fig:twopha_pcahexx1701}, imply that the spectral fitting results
for \object{XTE J1701-462} should also apply to the Sco-like Z
sources. 

In stage IV, the HB and NB are no longer observed, and the CDs/HIDs
resemble those of the bright persistent GX atoll sources \object{GX
9+1},\object{GX 9+9}, and \object{GX 3+1}. These sources are believed
to accrete at accretion rates between those of the classical atoll
sources (like \object{Aql X-1} and \object{4U 1608-52}) and those of
the Z sources \citep{hava,va2006}. Based on patterns traced out in the
CDs, they are very similar to \object{XTE J1701-462} in the Z stage
IV. We classify such patterns in the CD/HID for \object{XTE J1701-462}
into the NB/FB vertex and the FB, while the bright persistent atoll
sources have typical names ``lower banana'' and ``upper banana''
respectively \citep{hava,va2006}. It is possible that the lower banana
in the GX atoll sources is simply the NB/FB vertex and the upper
banana is the FB. This can be investigated with spectral fits, e.g.,
to see whether the upper banana shows evidence for typical FB spectral
evolution, as shown in Figure~\ref{fig:lum_Tbb_sub}.

In the brief atoll stage, \object{XTE J1701-462} resembles the
classical transient atoll sources \object{Aql X-1} and \object{4U
1608-52} from the CDs/HIDs and spectral fits (LRH07). The emission
size of the boundary layer remains constant across the states, and the
inner disk radius remains constant in the SS. The increase of the
Comptonization in the HS is also accompanied by an increase in
continuum power, i.e., the integrated rms in the PDS. We also note
that \object{XTE J1701-462} in the atoll stage has a peak luminosity
of ${\sim}0.2$ $L_{\rm EDD}$, while \object{Aql X-1} and \object{4U
1608-52} show maxima of ${\sim}0.35$ $L_{\rm EDD}$ (LRH07). We note
that these values are scaling estimates only, as there might be large
corrections required to infer intrinsic emission for some components,
e.g., the BB which might be partially obscured.

\section{CONCLUSIONS}
\label{sec:conclusion}
Our results offer a major departure from the classical view of Z
sources. The Sco-like Z track is traced out at nearly constant
$\dot{m}$, while the three branches are tied to different physical
mechanisms that function like instabilities tied to the more stable Z
vertices. On the other hand, the secular changes, which are driven by
variations in $\dot{m}$, unite, in sequence, all of the subclasses of
atoll and Z sources.

The conclusions 3--6 below are primarily derived from our chosen
spectral model, especially from the MCD component. The behavior of the
BB component is more difficult to understand, as this component might
be significantly affected by mass loss, obscuration, etc. We
acknowledge the need to test this model on the persistent Z sources to
determine whether such conclusions can be generalized. We also
acknowledge the need to further consider systematic problems
associated with the model and with the literal interpretation of
physical quantities derived from the spectral parameters. Finally,
there are alternative spectral models that convey different physical
interpretations for the mechanisms of Z branches
\citep[e.g.,][]{chbaja2008}, and detailed comparisons are required for
these models and their associated predictions.

\begin{enumerate}
\item In the nearly-20-month-long outburst in 2006--2007, \object{XTE
J1701-462} evolves through the characteristic behaviors of the
Cyg-like Z, Sco-like Z, and atoll sources as its luminosity decreases
from super-Eddington values toward quiescence. Our spectral fits
suggest that as $\dot{m}$ decreases NS LMXBs change from Cyg-like Z,
via Sco-like Z, to atoll sources.

\item As the $\dot{m}$ decreases, the HB disappears first, followed by
the NB, and finally by the FB. Despite the substantial secular changes
during the outburst, the HB/NB and NB/FB vertices trace out two
distinct lines in the HID. With the disappearance of the FB, the NB/FB
vertex smoothly evolves into the atoll track traced out at the lowest
accretion rate, beginning at $L_{\rm X}\sim 0.2$ $L_{\rm EDD}$. The
full length of the NB shortens with the decrease in luminosity.

\item In the atoll SS, the disk maintains a constant inner radius, at
a value presumed to match the ISCO, and the spectral evolution follows
$L_{\rm MCD} \propto T_{\rm MCD}^{4}$. Deviations from this behavior,
in the form of increasing $R_{\rm MCD}$ with $\dot{m}$, are found in
the NB/FB (lower) vertex during the Z-source stages. The truncated
disk at larger radius is attributed to the effect of reaching the
local Eddington limit at the inner disk radius.

\item The Sco-like FB is traced out when the disk shrinks back toward
the atoll stage value at constant $\dot{m}$. This appears to be an
instability in which the disk temporarily moves to reverse the
truncation level set by the NB/FB vertex. The Cyg-like FB, which is of
the ``dipping'' type in this source, cannot be satisfactorily fit with our
spectral model, and its nature is unknown.

\item As the source evolves along the NB from the upper to lower
vertices, the main spectral variation is the apparent size of the BB
emission. This might be due to additional matter supplied to the
boundary layer via the onset of a radial flow. Alternatively, there
might be geometric effects that alter our view of the boundary layer
while the source transverses the NB at constant $\dot{m}$. The NB
seems to represent another type of instability in Z sources, since
spectral evolution is faster on the NB than in its two connecting
vertices.

\item The Sco-like HB is traced out when some of the energy in the
disk is converted into a hard X-ray component, presumably via
Comptonization, while the disk $\dot{m}$ remains roughly
constant. Increasing continuum power in the PDS is also detected. The
Cyg-like HB is much longer than the Sco-like one. Its upturn resembles
the Sco-like HB in intensity range, strong Comptonization, and
continuum power. Thus the upturn of the Cyg-like HB probably has the
same nature as the Sco-like HB. The non-upturn part of the Cyg-like HB
lies along the same line traced by the HB/NB vertex in the HID. Only
weak Comptonization is detected, but there is also strong continuum
power, making this particular portion of the Cyg-like HB to appear
unique and puzzling.

\item Finally we speculate as to how these results can tie into
theoretical investigations of accretion disks at high luminosity. Like
the lower vertex, the upper vertex appears to be a more stable source
condition than the Z branches. The upper vertex is more commonly seen
when the source is at the highest levels of luminosity, and it is
associated with an increased efficiency in the disk for passing matter
through to the boundary layer. These same properties distinguish the
slim disk model from the standard thin disk. We therefore hypothesize
that the two vertices coincide with the two disk models and that
evolution up the NB represents the transition to the slim disk. The HB
association with a stronger jet can then be interpreted as an apparent
requirement that the slim disk be in place before the jet is able to
attain higher luminosity, which is coupled to the appearance of
increased Comptonization.

\end{enumerate}

\end{document}